\documentclass[aps,prd,twocolumn,superscriptaddress,nofootinbib,preprintnumbers]{revtex4-1}
\usepackage{graphics} 
\usepackage{epsfig}
\usepackage[usenames]{color}
\usepackage{verbatim}
\usepackage{amsmath,amsfonts,amssymb}

\def\pD  {p_{_D}}
\def\eD  {e_{_D}}
\def\pDbar  {\bar{p}_{_D}}
\def\eDbar  {\bar{e}_{_D}}
\def\HD  {H_{_D}}
\def\gammaD  {\gamma_{_D}}
\def\AD  {A_{\rm D}}

\def\muD {\mu_{_{D}}}
\def\mpD {m_{\rm p}}
\def\meD {m_{\rm e}}
\def\mHD {m_{_{\rm H}}}
\def\Mg  {M_\gamma}

\def\mpDsym {m_{\rm p, sym}}

\def\npD {n_{\rm p}}
\def\neD {n_{\rm e}}
\def\nHD {n_{_{\rm H}}}
\def\nDM {n_{\rm DM}}

\def\UD   {{\rm U}_{\rm D}(1)}
\def\aD   {\alpha_{_D}}
\def\aDuni{\alpha_{D, \rm uni}}
\def\aDsym{\alpha_{D, \rm sym}}
\def\aDsymT{\tilde{\alpha}_{D, \rm sym}}
\def\xD   {x_{_{D}}}
\def\vrel {v_{\rm rel}}
\def\BSF  {_{_{\rm BSF}}}
\def\FO   {_{_{\rm FO}}}
\def\DR   {_{_{\rm DR}}}
\def\DM   {_{_{\rm DM}}}

\def\cm  {{\rm cm}}
\def\MeV {{\rm MeV}}
\def\GeV {{\rm GeV}}
\def\TeV {{\rm TeV}}

\newcommand{\slashed}[1]{{#1}\hspace{-2mm}/}

\begin{document}

\title{Signals from dark atom formation in halos}

\author{Lauren Pearce}
\affiliation{William I. Fine Theoretical Physics Institute, School of Physics and Astronomy, University of Minnesota, Minneapolis, MN 55455 USA}

\author{Kalliopi Petraki}
\affiliation{Nikhef, Science Park 105, 1098 XG Amsterdam, The Netherlands}

\author{Alexander Kusenko}
\affiliation{Department of Physics and Astronomy, University of California, Los Angeles, CA 90095-1547, USA}
\affiliation{Kavli Institute for the Physics and Mathematics of the Universe (WPI), University of Tokyo, Kashiwa, Chiba 277-8568, Japan}

\preprint{NIKHEF-2015-003}
\preprint{FTPI-MINN-15-05}

\begin{abstract}
We consider indirect detection signals of atomic dark matter, with a massive dark photon which mixes kinetically with hypercharge. In significant regions of parameter space, dark matter remains at least partially ionized today, and dark atom formation can occur efficiently in dense regions, such as the centers of galactic halos. The formation of dark atoms is accompanied by emission of a dark photon, which can subsequently decay into Standard Model particles. We discuss the expected signal strength and compare it to that of annihilating dark matter. As a case study, we explore the possibility that dark atom formation can account for the observed 511 keV line and outline the relevant parameter space.
\end{abstract}

\maketitle

\section{Introduction}

Atomic dark matter arises in models in which dark matter (DM) couples to a dark Abelian gauge force, and the relic DM abundance is due to a dark particle-antiparticle asymmetry (see e.g.~\cite{Kaplan:2009de,Kaplan:2011yj,Petraki:2011mv,vonHarling:2012yn}). It is a minimal implementation of the asymmetric DM scenario, which is motivated by the similarity of the observed dark and ordinary matter abundances~\cite{Davoudiasl:2012uw,Petraki:2013wwa,Zurek:2013wia,Boucenna:2013wba}. Moreover, atomic DM is a simple possibility for self-interacting DM. The latter has emerged as an alternative to the collisionless cold DM paradigm that can resolve the discrepancies between collisionless cold DM simulations and observations of the galactic structure~\cite{Spergel:1999mh,Wandelt:2000ad,Faraggi:2000pv,Kusenko:2001vu,Mohapatra:2001sx,
Kaplan:2009de,Feng:2009mn,Feng:2009hw,Loeb:2010gj,Alves:2010dd,Kaplan:2011yj,Rocha:2012jg,Peter:2012jh,
Vogelsberger:2012ku,Vogelsberger:2012sa,Zavala:2012us,Cline:2012is,Cyr-Racine:2013fsa,CyrRacine:2012fz,
Cline:2013pca,Petraki:2014uza,Foot:2014uba,Foot:2014mia,Tulin:2013teo,Cline:2013zca,Kouvaris:2013gya,
Boddy:2014yra,Hochberg:2014dra,Kouvaris:2014uoa}.\footnote{For an overview of the collisionless cold DM challenges at galactic and subgalactic scales, see Ref.~\cite{Weinberg:2013aya}.}
In fact, because in the atomic DM scenario the dark particles couple to a light force mediator, the DM self-interactions may manifest as long-range, with scattering cross-sections which decrease with increasing velocity. This feature implies that the DM self-interactions may affect significantly the dynamics of smaller haloes, while having negligible impact on larger scales~\cite{CyrRacine:2012fz,Cline:2013pca,Petraki:2014uza,Foot:2014uba}, where the collisionless cold DM paradigm has been very successful. However, there is also plenty of parameter space in which atomic DM behaves as collisionless.

It has been previously observed that asymmetric DM with long-range self-interactions may produce detectable indirect detection signals due to the formation of DM bound states inside haloes today~\cite{Pearce:2013ola}. Dark bound-state formation is accompanied by emission of a dark force mediator, which can subsequently decay into Standard Model (SM) particles, via a so-called portal interaction. In this work, we explore this possibility within the scenario of atomic DM coupled to a light but massive dark photon which kinetically mixes with the hypercharge.

The atomic character of DM is evident in the case of asymmetric DM coupled to a massless dark photon; in Ref.~\cite{Petraki:2014uza}, it was rigorously shown that DM remains atomic even if the dark U(1) force is mildly broken. In the case of a massless dark photon, gauge invariance mandates that DM must be multicomponent: The dark asymmetry generation has to occur via gauge-invariant operators, therefore the net gauge charge carried by the asymmetric population of a dark species must be compensated by an opposite gauge charge carried by an asymmetric population of another dark species. The relic abundances of the two dark species may (partially) combine into dark atoms in the early universe, as well as inside haloes today. This is, of course, analogous to ordinary protons and electrons, whose electric charges compensate each other. 

Reference~\cite{Petraki:2014uza} examined in detail the case of the dark photon acquiring mass via a dark Higgs mechanism in the early universe. If the dark photon mass is small, the dark phase transition occurs cosmologically at late times, thus leaving unaffected the preceding cosmology, including the dark asymmetry generation and the dark recombination. It then follows that -- as in the case of a massless dark photon -- DM has to be multicomponent, with the dark ions potentially forming dark atoms in the early universe.  The range of dark photon masses for which DM is atomic includes much of the parameter space where the DM self-interaction is sufficiently strong to affect the dynamics of haloes; thus, the multicomponent and atomic character of DM may not be neglected when studying this effect~\cite{Petraki:2014uza}. Moreover, the rich composition of DM can result in indirect detection signals, some of which we explore in this work. In particular, dark atomic bound states may form today from the residual ionized fraction of DM. The dark photons emitted in these processes -- being massive -- may subsequently decay into SM particles via their kinetic mixing with hypercharge.

Related ideas, involving indirect detection signals from excitations and de-excitations of bound states of asymmetric DM in haloes, have been explored in Refs.~\cite{Frandsen:2014lfa,Cline:2014eaa,Boddy:2014qxa,Detmold:2014qqa}. Indirect signals from the formation of bound states by (weakly-interacting) symmetric DM have been discussed in Refs.~\cite{Pospelov:2008jd,MarchRussell:2008tu,Shepherd:2009sa}.

This paper is organised as follows. In Sec.~\ref{sec:Model_Introduction}, we briefly review the atomic DM scenario. 
In Sec.~\ref{sec:Signal}, we discuss the signal expected from the formation of dark atoms in haloes and we compare it to that of annihilating symmetric DM. In Sec.~\ref{sec:511 keV}, we consider the possibility of the 511 keV line being produced from the decays of dark photons emitted during the formation of dark atoms in the Milky Way. 
We conclude in Sec.~\ref{sec:Conc}.

\section{Atomic dark matter}
\label{sec:Model_Introduction}

We consider fermionic DM charged under a dark $\UD$ gauge symmetry, under which the SM fermions are uncharged. We also assume that the low-energy theory describing the dark sector possesses a global U(1) symmetry -- a dark baryon number -- under which DM transforms, and that the relic population of DM is due to an excess of dark baryons over dark antibaryons. As discussed in the introduction (see~\cite{Petraki:2014uza} for details), if the $\UD$ symmetry is unbroken or mildly broken, gauge invariance implies that DM consists of two stable particle species, oppositely charged under $\UD$; we shall call them the dark proton, $\pD$, and the dark electron, $\eD$, and assume they are fundamental. We denote their masses by $\mpD$ and $\meD$, and take their $\UD$ charges to be $q_{\rm p} = +1$ and $q_{\rm e} = -1$ respectively.  We also take $\mpD \geqslant \meD$. 
The low-energy physics of interest is governed by the Lagrangian
\begin{align}
{\cal L} 
&=  \pDbar(i \slashed{D} - \mpD) \pD + \eDbar(i \slashed{D} - \meD) \eD 
\nonumber \\
&- \frac{1}{4} F_{\rm D \, \mu\nu}^{} F_{\rm D}^{\mu\nu} 
+ \frac{1}{2} \Mg^2 \, {\AD^{}}_\mu \AD^\mu
\ ,
\label{eq:L}
\end{align}
where $F_{\rm D}^{\mu\nu} = \partial^\mu \AD^\nu - \partial^\nu \AD^\mu$, with $\AD$ being the dark-photon field.  (As in QED, we use $\AD^\mu$ for the field in the Lagrangian, and $\gammaD$ for the dark photon when discussing processes in which it participates.) $\Mg$ is the dark photon mass, which may be generated either via the Higgs mechanism or the St\"{u}ckelberg mechanism. The covariant derivative for $\pD$ and $\eD$ is $D^\mu  = \partial^\mu + i q_j g \AD^\mu$, where $q_j$ is the respective charge and $g$ is the gauge coupling of the dark force.  In the following, we more commonly use the dark fine structure constant $\aD \equiv g^2/4\pi$.

Atomic bound states of dark protons and dark electrons exist if $\Mg < \aD \muD$, where $\muD$ is the $\pD-\eD$ reduced mass, 
\begin{equation} 
\muD \equiv \frac{\mpD \meD}{\mpD+\meD} \, .
\label{eq:muD}
\end{equation} 
The ground state of the dark Hydrogen atom $\HD$, has mass
\begin{equation}
\mHD = \mpD + \meD - \Delta  \, ,
\label{eq:mH}
\end{equation} 
where $\Delta$ is the ground-state binding energy, 
\begin{equation}
\Delta \simeq \dfrac{\aD^2 \muD}{2} \left( 1 - \dfrac{M_\gamma}{\aD \muD} \right)^2.
\label{eq:Delta}
\end{equation}
Note that $\muD$ and $\mHD$ satisfy the consistency relation
\begin{equation}
4 \muD \leqslant \mHD + \Delta \, ,
\label{eq:consistency}
\end{equation}
where the equality is realized for $\mpD = \meD$. (In most cases, it suffices to approximate \eqref{eq:consistency} with $4\muD \lesssim \mHD$.)

Provided that $M_\gamma < \Delta$, dark atoms can form radiatively, with emission of a dark photon,
\begin{equation}
\pD + \eD \to \HD +\gammaD \, .
\label{eq:BSF}
\end{equation}
Evidently, the upper limit on $\Mg$ for the radiative formation of dark atoms is stronger than that for the existence of dark atomic bound states. Dark protons and dark electrons bind partially into dark Hydrogen atoms during the epoch of dark recombination in the early universe, which has been studied in detail in Ref.~\cite{CyrRacine:2012fz}. The phenomenology of DM today is largely determined by the residual ionization fraction,
\begin{equation}
\xD \equiv \frac{\npD}{\nHD+\npD} \, ,
\label{eq:xD def}
\end{equation}
where $n_j$ denotes the number density of the $j$ dark element today. Obviously, $\npD = \neD$. In the following, we will often denote $\nHD + \npD = \nDM$. The residual ionization fraction can be approximated by~\cite{CyrRacine:2012fz} 
\begin{equation}
\xD \approx \min \left[ 1, \: 10^{-10} \ \dfrac{\xi\DR}{\aD^4} \left( \dfrac{\mHD \muD}{\mathrm{GeV}^2} \dfrac{1}{PS} \right) \right] \, .
\label{eq:xD}
\end{equation}
In Eq.~\eqref{eq:xD} and in the following, the parameter  $\xi = \min[1,T_D \slash T_V]$ is determined by the ratio of the dark sector temperature $T_D$, to the ordinary sector temperature $T_V$. $\xi$ may in general vary with time, albeit typically mildly. The subscript denotes the relevant time,  with ``DR"  referring to the epoch of dark recombination.  In Eq.~\eqref{eq:xD} we have inserted a phase-space suppression factor relevant when $\Delta \sim M_\gamma$,
\begin{equation}
PS = \sqrt{1 - \dfrac{M_\gamma^2}{(\Delta + \muD \vrel^2 \slash 2)^2} } \, ,
\end{equation}
where $\vrel$ is the average relative velocity of $\pD$ and $\eD$ at the relevant time (for Eq.~\eqref{eq:xD}, this is the time of dark recombination). 
Although the approximation of Eq.~\eqref{eq:xD} for the ionization fraction works well when $\xD = 1$ and $\xD \ll 1$, it is less satisfactory when $\xD \lesssim 1$.  However, as the ionization fraction depends strongly on $\aD$, this is a relatively small region of parameter space.

The efficient annihilation of the dark antiparticles ($\pDbar$ and $\eDbar$) in the early universe, sets a lower limit on the DM annihilation cross-section, and therefore on $\aD$. In the presence of a particle-antiparticle asymmetry, the dark antiparticles are diminished to less than 10\% (1\%) of the DM density, if the annihilation cross-section is only about 1.4 (2.4) times or more the value required for symmetric thermal relic DM~\cite{Graesser:2011wi}. For DM consisting of Dirac fermions coupled to a dark photon, the coupling required to obtain the observed DM density in the symmetric regime has been calculated in Ref.~\cite{vonHarling:2014kha}, taking into account both the Sommerfeld enhancement of the DM annihilation, as well as the formation and decay of dark particle-antiparticle bound states, which also contributes to the overall DM destruction rate. Moreover, in the atomic DM scenario, $\pD\pDbar$ can annihilate either into $\gammaD\gammaD$ or $\eD\eDbar$, with equal cross-sections; this doubles the total annihilation rate. Thus, as the minimum value of $\aD$ for efficient annihilation, it suffices to consider that determined for symmetric DM in the presence of the $\gammaD\gammaD$ annihilation channel only. We require $\aD \gtrsim \aDsym  (\mpD)$, with~\cite{vonHarling:2014kha}
\begin{multline}
\frac{\aDsym (\mpD)}{ 0.031 \ \xi\FO^{1/2} }  \simeq
\left(\frac{\mpD}{\TeV}\right) \!
\left[1 + \left( \frac{\mpD}{\TeV} \right)^{1.28} \right]^{-0.328} ,
\label{eq:alpha ann}
\end{multline}
where the subscript ``FO" refers to the time of the DM freeze-out. Since $\meD \leqslant \mpD$, this obviously also ensures efficient annihilation of $\eDbar$. In the next section, we shall see that this minimum value of $\aD$ implies both upper and lower bounds on the signal strength expected from the radiative formation of dark atoms today.

\section{Signal production \label{sec:Signal}}

\subsection{Dark atom formation \label{sec:HD formation}}

The cross-section times relative velocity, for the radiative formation of dark atoms shown in \eqref{eq:BSF} is~\cite{AkhiezerMerenkov_sigmaHydrogen}
\begin{align}
(\sigma \vrel)\BSF &\simeq 
\dfrac{2^9 \pi^2 \, \aD^2}{3 \, \muD^2}  
\: \dfrac{\zeta^5}{(1 + \zeta^2)^2} 
\: \dfrac{e^{-4 \zeta {\rm arccot}(\zeta)}}{1 - e^{-2 \pi \zeta}} 
\cdot PS \, ,
\label{eq:sigmav_full}
\end{align}
where $\zeta$ is the ratio of the Bohr momentum to the initial state center of mass momentum, 
\begin{equation}
\zeta = \dfrac{\mu \aD}{\mu \vrel} = \dfrac{\aD}{\vrel},
\end{equation}
and again we have introduced a phase-space suppression factor. For the indirect detection signals produced from dark atom formation inside haloes, we will be mostly interested in the regime  $\vrel \ll \aD$. In this limit,
\begin{multline}
(\sigma \vrel)\BSF \approx 3.5 \times 10^{-21} \ \frac{\rm cm^3}{\rm s}
\\
\times
\left( \frac{\aD}{10^{-2}} \right)^3
\left( \frac{10~\rm GeV}{\muD}  \right)^2
\left( \dfrac{10^{-3}}{\vrel} \right)
\cdot PS \, ,
\label{eq:sigmav_approx}
\end{multline}
which exhibits the familiar $\propto 1/\vrel$ scaling of Sommerfeld-enhanced processes at low velocities. This low-velocity enhancement saturates when the momentum transfer between the incoming particles becomes less than the mediator mass, i.e.~at $\vrel \lesssim \Mg/\muD$; for a massive dark photon, in Eq.~\eqref{eq:sigmav_approx}, we may thus replace $\vrel \to \max[\vrel, \Mg/\muD]$.

Although we account for the saturation of the cross section, we do not consider here resonances which may occur due to the nonzero mediator mass.  Such resonances affect only a small region of parameter space, and serve to enhance the signal.

Being Sommerefeld enhanced, the dark atom formation becomes very efficient in the non-relativistic environment of DM haloes. However, the expected signal today depends on the interplay between the strength of the interaction cross-section given above, and the early universe cosmology, which determines the residual ionization fraction of DM. We discuss this interplay in Sec.~\ref{sec:signal strength}.

Partial wave unitarity sets an upper limit on the total inelastic cross-section; in the non-relativistic regime, this is  $\sigma_{{\rm inel},j} \leqslant (2j+1)\pi/(\muD^2 \vrel^2)$~\cite{Griest:1986yu}. This gives a rough estimate of the range of validity of Eq.~\eqref{eq:sigmav_full},\footnote{A more precise determination of $\aDuni$ requires partial wave decomposition of the differential cross-section leading to Eq.~\eqref{eq:sigmav_full}, which is beyond the scope of the present work. Such precision is not significant for our purposes.}
\begin{equation}
\aD \lesssim \aDuni \approx 0.5 \, .
\label{eq:alpha uni}
\end{equation}

\subsection{Dark photon decay \label{sec:gammaD decay}}

We have noted that the SM particles are uncharged under the $\mathrm{U}_\mathrm{D}(1)$ symmetry, while the dark fermions are uncharged under the SM gauge group. However, the dark sector and the SM sector may couple through kinetic mixing of the U(1) gauge bosons.  We introduce the renormalizable operator~\cite{Holdom:1985ag,Foot:1991kb}
\begin{equation}
\mathcal{L}_\mathrm{mix} = \dfrac{\epsilon}{2} F_{\mathrm{Y}\mu \nu} F_{\mathrm{D}}^{\mu \nu}.
\label{eq:kinetic mix}
\end{equation}
Once $\mathrm{SU}_\mathrm{L}(2)$ is broken, this term induces a coupling between the dark photon and the SM photon, as well as between the dark photon and the $Z$ boson.  

The dark photon may decay into charged fermions, at the rate~\cite{Batell:2009yf}
\begin{equation}
\Gamma_{\gammaD \rightarrow f^+ f^-} = f_\mathrm{EM} \cdot \dfrac{\epsilon^2}{3} \alpha_\mathrm{EM} M_\gamma \, ,
\label{eq:gammaD to ff}
\end{equation}
where $\alpha_{\rm EM} \simeq 1/137$ is the electromagnetic fine-structure constant, and $f_\mathrm{EM}$ accounts for the number of kinematically allowed decay channels.  The dark photon may also decay into $\bar{\nu} \nu$ via its mixing with the $Z$ boson, with rate
\begin{equation}
\Gamma_{\gammaD \rightarrow \bar{\nu} \nu} = \dfrac{\epsilon^2}{3} \alpha_\mathrm{EM} M_\gamma \cdot \dfrac{3 M_\gamma^4}{4 \cos^2(\theta_W) m_Z^4} \, .
\label{eq:gammaD to 2nu}
\end{equation}
Although the decay $\gammaD \rightarrow \gamma \gamma$ is forbidden, the dark photon can decay to three photons via a charged fermion loop, at the rate~\cite{Pospelov:2008jk}
\begin{equation}
\Gamma_{\gammaD \rightarrow 3\gamma} = \dfrac{17 \epsilon^2 \alpha_\mathrm{EM}^4 M_\gamma^9}{2^7 3^6 5^3 \pi^2 \tilde{m}_e^8},
\label{eq:gammaD to 3gamma}
\end{equation}
where $\tilde{m}_e = 511$~keV is the ordinary electron mass.  The decay to charged fermions dominates unless it is kinematically forbidden.

The cosmology of relic dark photons was considered in~\cite{Petraki:2014uza}.  Provided that the $\UD$-breaking phase transition occurs before BBN, dark photons with a mass above 1.022 MeV decay before BBN if
\begin{equation}
\epsilon > \dfrac{10^{-10}}{f_\mathrm{EM}^{1 \slash 2}} 
\left( \dfrac{10 \; \mathrm{MeV}}{M_\gamma} \right)^{1 \slash 2}. 
\label{eq:epsilon BBN}
\end{equation}
Otherwise, we must require that during BBN $\xi_\mathrm{BBN} = T_D \slash T_V < 0.6$, to satisfy constraints on relativistic energy density.  In this case, further constraints (such as avoiding altering the time of matter-radiation equality) may be relevant.  Experimental bounds on $\epsilon$ are explored in \cite{Jaeckel:2012yz,Lees:2014xha}.  For dark photon masses around $\sim$10~MeV, $\epsilon \lesssim 10^{-8}$ due to supernova 1987a, while for dark photon masses on the GeV scale, $\epsilon \lesssim 10^{-3}$ is constrained by collider experiments.

Therefore, the relic dark photons will typically have decayed before BBN, and for an even wider range of mixings $\epsilon$, before the present day.  The indirect detection signals relevant today must arise from ongoing production of dark photons.  Consequently, we focus on dark photon emission which accompanies ongoing dark atom formation in DM haloes.

\subsection{Signal strength \label{sec:signal strength}}

Bound states are produced at the rate per volume
\begin{equation}
\frac{d\Gamma\BSF}{dV} = (\sigma \vrel)\BSF \npD \neD =\frac{\xD^2 (\sigma \vrel)\BSF}{\mHD^2} \: \rho\DM^2 \, ,
\label{eq:BSF_rate}
\end{equation}
where $\rho\DM$ is the DM density, and we took into account that 
\begin{align}
\rho_\mathrm{DM} &= \nHD \, \mHD + \npD \, \mpD + \neD \, \meD \nonumber \\
&= (1-\xD) \nDM \mHD + \xD \, n_\mathrm{DM} \, (\mpD + \meD)  \nonumber \\
&\simeq (1-\xD) \nDM \mHD + \xD \, n_\mathrm{DM} \, \mHD  \nonumber \\
&= \nDM \, \mHD \, , \nonumber 
\end{align}
with $\nDM = \nHD + \npD$, and the binding energy was neglected. This allowed us to substitute $\npD\neD = \xD^2 \nDM^2 = \xD^2 \rho\DM^2 \slash \mHD^2$.

Evidently, the signals produced by dark atom formation scale with the DM density in the same way as DM annihilation. 
We define the signal strength $s\BSF$, as\footnote{More properly, $s\BSF$ should be defined in terms of the average $(\sigma \vrel)\BSF$ over the velocity distributions of the participating particles. For $\vrel \ll \aD$, which is the regime of greatest interest, $(\sigma \vrel)\BSF \propto 1/\vrel$, and for a Maxwellian distribution, $\langle1/\vrel\rangle = \sqrt{6/\pi} / \bar{v}_{\rm rel} \simeq 1.38 / \bar{v}_{\rm rel}$, where $\bar{v}_{\rm rel}$ is the rms value of the relative velocity. However, for simplicity, we will use $\langle1/\vrel\rangle \to 1/\bar{v}_{\rm rel}$, and denote $\bar{v}_{\rm rel}$ with $\vrel$. A proper averaging would enhance the expected signal.}
\begin{equation}
s\BSF \equiv  \dfrac{\xD^2 (\sigma \vrel)\BSF}{\mHD^2} \, ,
\label{eq:sBSF}
\end{equation}
which facilitates the comparison with signals expected from annihilating DM, and contains the combination of parameters that is generally constrained by observations. In the regime $\aD \gg \vrel$, 
\begin{align}
s\BSF \approx \mathrm{min} \left[ \frac{2^9 \pi^2 \, \aD^3 \times PS}{3 e^4 \, \mHD^2 \muD^2 \vrel}, 
\frac{3 \cdot 10^{-19}~\GeV^{-4} \: \xi\DR^2}{\aD^5 \vrel \times PS} 
\right] , 
\label{eq:sBSF approx}
\end{align}
where inside the square brackets, the factor on the left corresponds to the fully ionized case and the factor on right corresponds to the partially ionized regime. The energy available to the dark photons emitted in the formation of dark atoms is the sum of the binding energy and the kinetic energy of the reduced system,
\begin{equation}
\omega_\gamma = \frac{1}{2} \muD (\aD^2 +\vrel^2)  \, .
\label{omega}
\end{equation} 
In most cases of interest, $\vrel \ll \aD$ and $\omega_\gamma \simeq \Delta$.

\begin{figure}
\includegraphics[scale=.6]{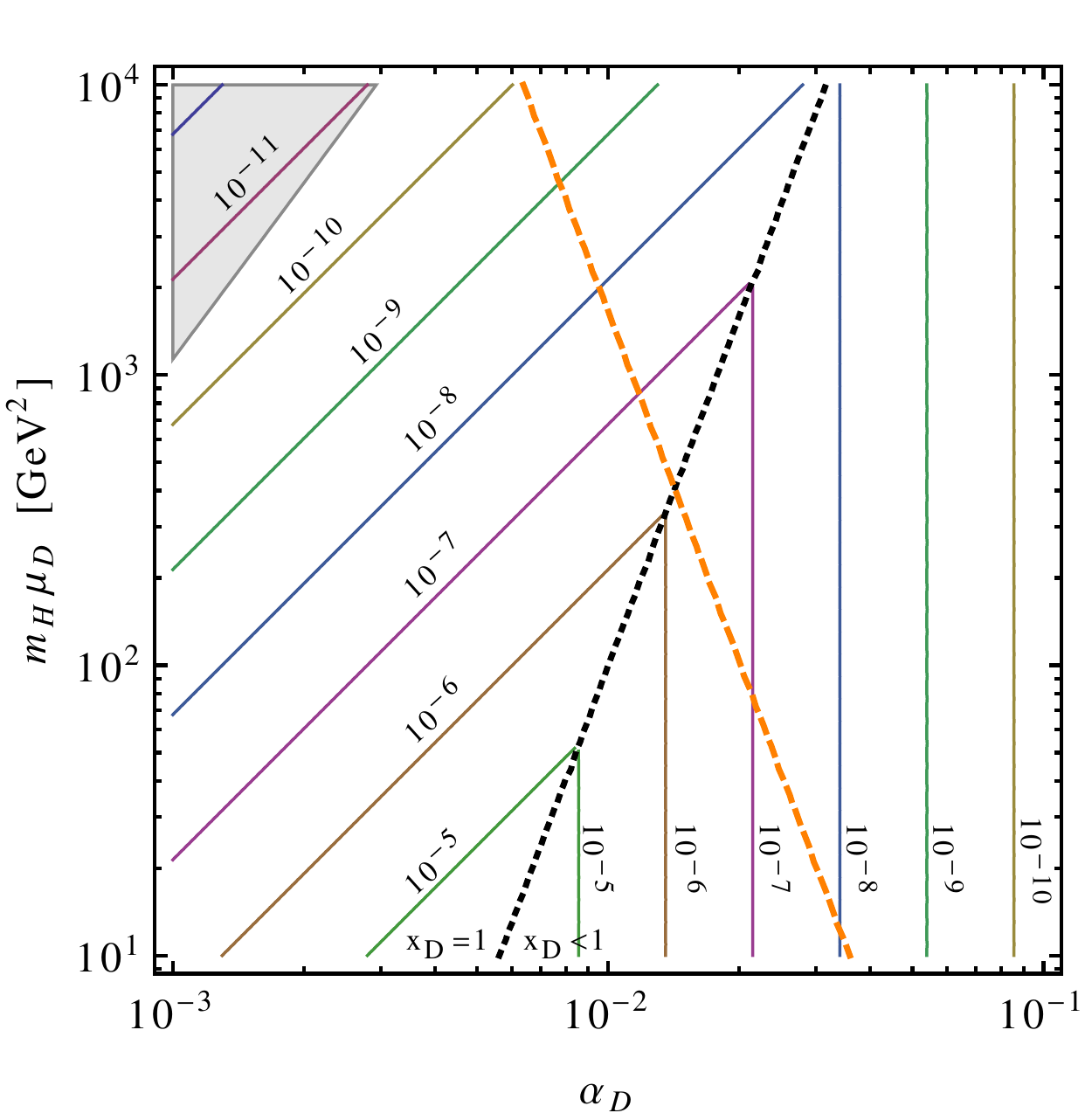}
\caption{Countour plot of the signal strength $s_{_{\rm BSF}}$, from bound state formation, in units of ${\rm GeV}^{-4}$. (For convenience, we note that ${\rm GeV}^{-4} \simeq 10^{-11} {\rm cm^3 \, s^{-1} \, TeV^{-2}}$.) We have used $\xi_{_{\rm DR}} = 1$, $v_{\rm rel} = 200 \; {\rm km/s}$. 
We assume that the dark photon mass is less than the binding energy, $M_\gamma < \Delta$, such that the radiative formation of bound states is possible. To the left of the orange dashed line, the decay $\gamma_{_D} \rightarrow e^+ e^-$ is forbidden, because the binding energy is $\Delta \approx \alpha_{_D}^2 \mu_{_D} \slash 2 < 1.022~{\rm MeV}$ for all possible values of $\mu_{_D}$.  Above the orange dashed line, whether the decay $\gamma_{_D} \rightarrow e^+e^-$ is kinematically allowed depends on $M_\gamma$. In the grey shaded region, the coupling is insufficient to remove the symmetric DM component in the early universe, for any consistent choice of $m_{_{\rm H}}$, $\mu_{_D}$ (Eq.~\eqref{eq:alphaD range}).}
\label{fig:signal_1}
\end{figure}

A contour plot of the signal strength, $s\BSF$, is shown in Fig.~\ref{fig:signal_1}.  In the fully ionized regime, the signal strength increases with the coupling $\aD$; however, in the partially ionized regime, the signal strength decreases rapidly with increasing $\aD$.  This is a consequence of the dramatic decrease in the ionization fraction.  The sharp line at $\xD=1$ is a feature of the approximation of Eq.~\eqref{eq:xD}; an exact ionization fraction would round the sharp corners.  
The consistency relation of Eq.~\eqref{eq:consistency} implies that at every point in the ($\aD$, $\mHD\muD$) plane, there is a maximum value of the energy imparted into the dark photon which is attained for $\mpD = \meD$ and can be expressed as
\begin{equation}
\omega_\gamma \leqslant \omega_{\gamma, 0} \simeq \frac{1}{4} (\mHD \muD)^{1/2} \: (\aD^2 +\vrel^2) \, .
\label{eq:omega 0}
\end{equation} 
Since the radiative capture of $\pD, \eD$ to bound states is possible only for $\Mg < \omega_\gamma$, this condition constrains also the possible dark photon decay channels and the indirect detection signatures. 
In Fig.~\ref{fig:signal_1}, the orange dashed line marks $\omega_{\gamma,0} = 1.022~\MeV$; to the left of this line, the decay $\gammaD \rightarrow e^+ e^- $ is kinematically forbidden for any choice of $\Mg< \omega_{\gamma, 0}$. (However, the dark photon may still decay to neutrinos or three SM photons.) We see that if the dark photon decay into charged fermions is possible, the greatest signal strength that can be produced is $s\BSF \sim 10^{-6} \, \mathrm{GeV}^{-4}$. 
More generally, while the signal strength depends on various parameters and is not fully determined at a specific value of the dissipated energy, it is bounded both from above and below. We shall now explore the expected signal range.

\subsection{Circumscribing  the signal strength \label{sec:signal boundaries}}

Equations~\eqref{eq:alpha ann} and \eqref{eq:alpha uni} suggest that the DM coupling to the dark photons should be in the range
\begin{equation}
\aDsym(\mpD) < \aD < \aDuni   \, .
\label{eq:alphaD range}
\end{equation}
Moreover,
\begin{equation}
8 \Delta /\aD^2 = 4 \mu \lesssim \mHD \lesssim 2\mpD \, .
\label{eq:masses range}
\end{equation}
We shall now employ Eqs.~\eqref{eq:alphaD range} and \eqref{eq:masses range} to circumscribe the strength of the signal expected from dark atom formation in haloes today,  given by Eq.~\eqref{eq:sBSF}; in Sec.~\ref{sec:BSF-ann comparison} we compare it to the signal expected from annihilating symmetric DM. As we shall see, both the lower and the upper limit on $\aD$ given in \eqref{eq:alphaD range}, bound the expected signal strength from both above and below.

For convenience,  we first invert Eq.~\eqref{eq:alpha ann} numerically, to obtain the maximum dark proton mass for which annihilation is efficient in the early universe, given a coupling $\aD$, $\mpD \lesssim \mpDsym (\aD)$. We find the approximation
\begin{equation}
\frac{\mpDsym (\aD)}{\TeV} \approx \left(\frac{\aD}{0.031 \, \xi\FO^{1/2}}\right)
\left[1 + \left(\frac{\aD}{0.031 \, \xi\FO^{1/2}}\right)^{1.7}  \right]^{0.42} .
\label{eq:mp ann max}
\end{equation}
Furthermore, since $\mpDsym (\aD) \gtrsim \mpD \geqslant 2\muD = 4\Delta/\aD^2$, we may employ the above to obtain a lower bound on $\aD$ as a function of the binding energy, $\aD \gtrsim \aDsymT (\Delta)$. We find numerically
\begin{equation}
\frac{\aDsymT(\Delta)}{0.031 \, \xi\FO^{1/2}}  \approx 
\left(\frac{\Delta / \xi\FO}{240 \, \MeV}\right)^{\frac{1}{3}} 
\left[1 + \left(\frac{\Delta / \xi\FO}{240 \, \MeV}\right)^{0.5} \right]^{-0.13} .
\label{eq:alpha ann Delta}
\end{equation}

\subsubsection{Signal range at fixed $\aD$}

\begin{figure}
\includegraphics[scale=.6]{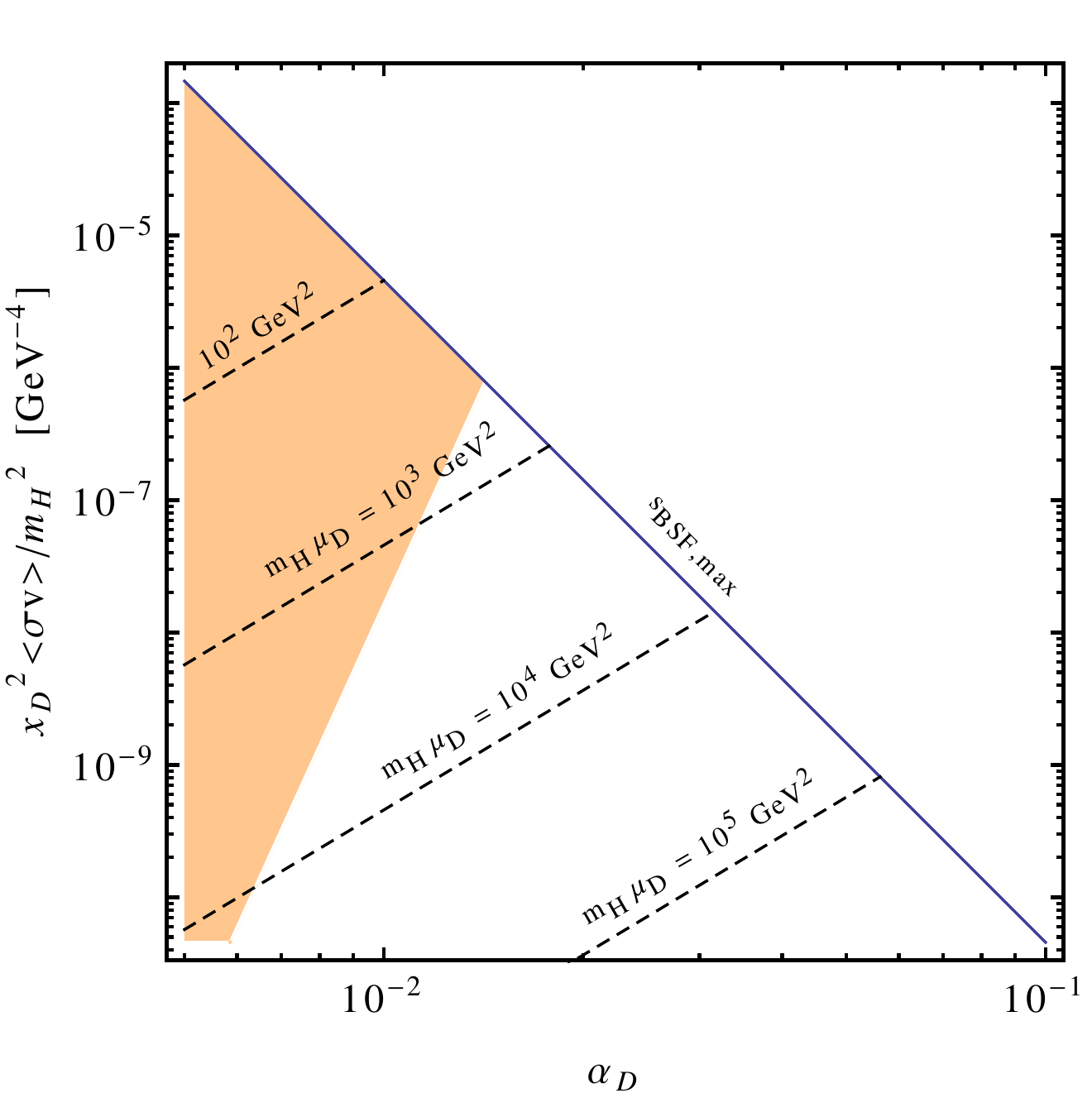}
\caption{Maximum signal strength $s_{_{\rm BSF}}$, for $\xi\DR = 1$, $\vrel = 200 \; \mathrm{km} \slash \mathrm{s}$.  We assume that the dark photon mass is always lower than the energy dissipated in the formation of bound states, such that the latter can occur radiatively. In the orange region, the decay $M_\gamma \rightarrow e^+ e^-$ is forbidden for all of parameter space.  The dashed lines correspond to the signal in the fully ionized region.
(Note that ${\rm GeV}^{-4} \simeq 10^{-11} {\rm cm^3 \, s^{-1} \, TeV^{-2}}$.)
}
\label{fig:signal_2}
\end{figure}

From Eq.~\eqref{eq:sBSF approx}, it follows that for a fixed value of $\aD$ there is a maximum signal strength that can be produced, which is given by
\begin{equation}
s_\mathrm{BSF,max} (\aD) \approx 
\frac{3\cdot 10^{-19}~\GeV^{-4} \ \xi\DR^2}{\aD^5 \, \vrel} \, ,
\label{eq:sBSFmax alpha}
\end{equation}
provided that $\aD\gg \vrel$ and $\Mg \lesssim \Delta$.  (For lower values of $\aD$, an expression for $s_\mathrm{BSF,max}$ can be obtained using Eq.~\eqref{eq:sigmav_full}.)  This maximum signal is produced if 
$\xi\DR (\mHD \muD) / \GeV^2 \lesssim 10^{10} \, \aD^4$ 
and is shown as a function of $\aD$ by the blue solid line in Fig.~\ref{fig:signal_2}. 

\begin{figure}
\includegraphics[width=0.9\linewidth]{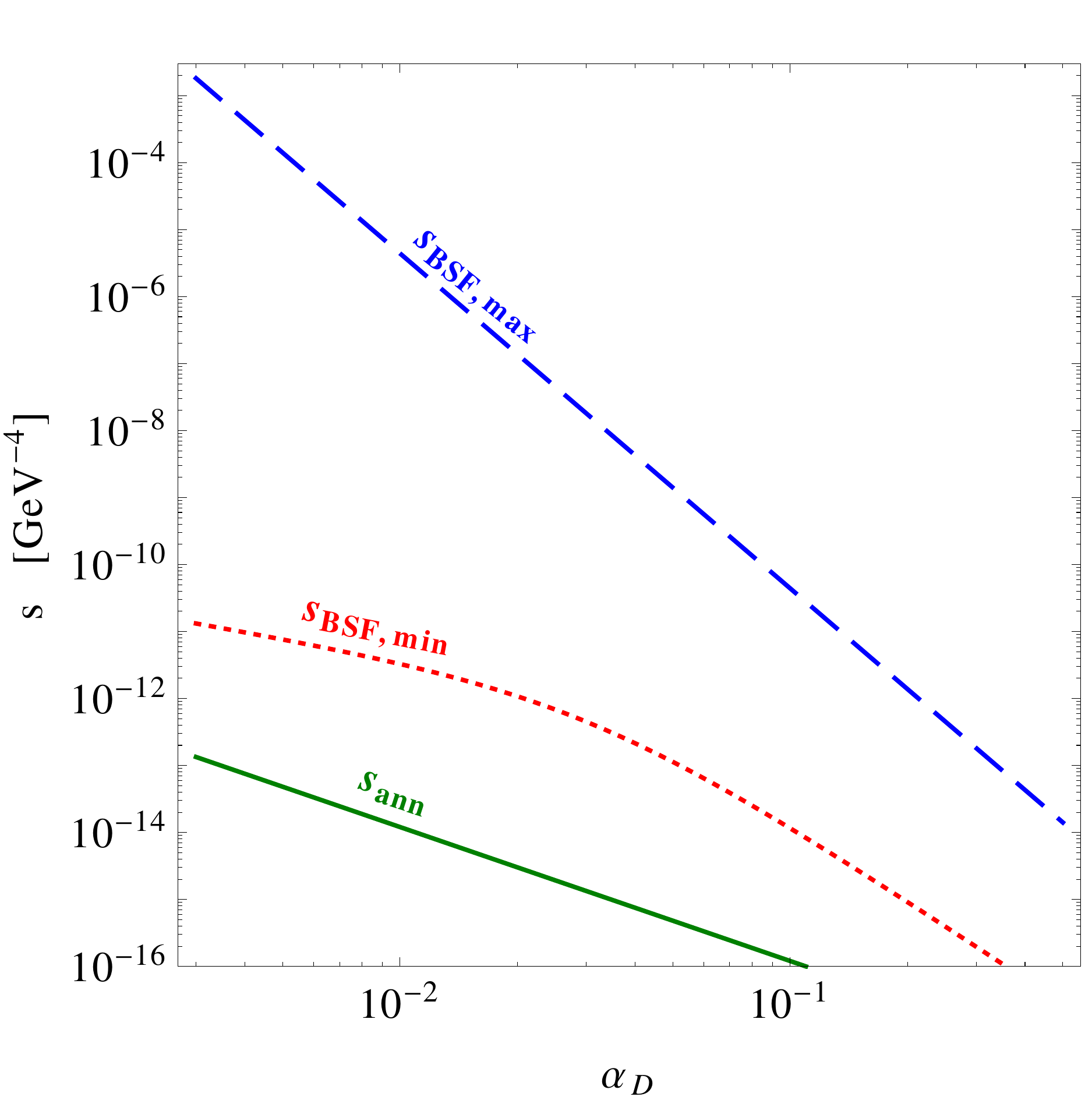}
\caption{Minimum (red dotted) and maximum (blue dashed) signal strength from dark atom formation, as a function of the dark fine structure constant, $\alpha_{_D}$.  The green solid line is the expected signal strength from the annihilation of symmetric DM whose coupling strength and mass are related by the observed DM density (see text for details). We have used $\xi_{_{\rm DR}} = \xi_{_{\rm FO}} = 1$ (which holds if the dark sector temperature is greater than the visible sector temperature at both times) and $v_{\rm rel} = 200~{\rm km/s}$ (smaller $v_{\rm rel}$ would enhance $s_{_{\rm BSF}}$).  For atomic DM, we assume that the dark photon mass is $0< M_\gamma < \Delta$, such that the radiative formation of bound states is possible and the decay of dark photons is kinematically allowed. (Note that ${\rm GeV}^{-4} \simeq 10^{-11} {\rm cm^3 \, s^{-1} \, TeV^{-2}}$.) }
\label{fig:comp_alpha}
\end{figure}

For larger values of $\mHD\muD$, the signal strength decreases; however, from Eqs.~\eqref{eq:masses range} and \eqref{eq:mp ann max} we find, $\mHD\muD \lesssim \mHD^2/4 \lesssim \mpD^2 \lesssim \mpDsym^2(\aD)$.
This implies a lower bound on $s\BSF$,
\begin{equation}
s_{\rm BSF,min} (\aD) \approx  
\frac{2^9 \pi^2}{3e^4} \frac{\aD^3}{\mpDsym^4(\aD) \ \vrel} \, .
\label{eq:sBSFmin alpha}
\end{equation}
In Fig.~\ref{fig:signal_2}, we sketch the signal strength $s\BSF$ as a function of $\aD$ for different values of $\mHD \muD$. In the orange region, the decay $\gammaD \rightarrow e^+e^-$ is forbidden for all choices of $\muD$ and $\mHD$ which satisfy the consistency relation~\eqref{eq:consistency}. In Fig.~\ref{fig:comp_alpha}, we plot both the maximum and the minimum signal strengths expected from bound state formation, over a larger parameter region.  We emphasize that for fixed $\aD$ the signal strength depends only on $\xi_\mathrm{DR}$; however, the minimum signal depends on $\xi_\mathrm{FO}$ as minimum coupling $\aD$ which permits sufficient annihilation of the symmetric component does.

\subsubsection{Signal range at fixed DM mass, $\mHD$}

\begin{figure}
\includegraphics[width=0.9\linewidth]{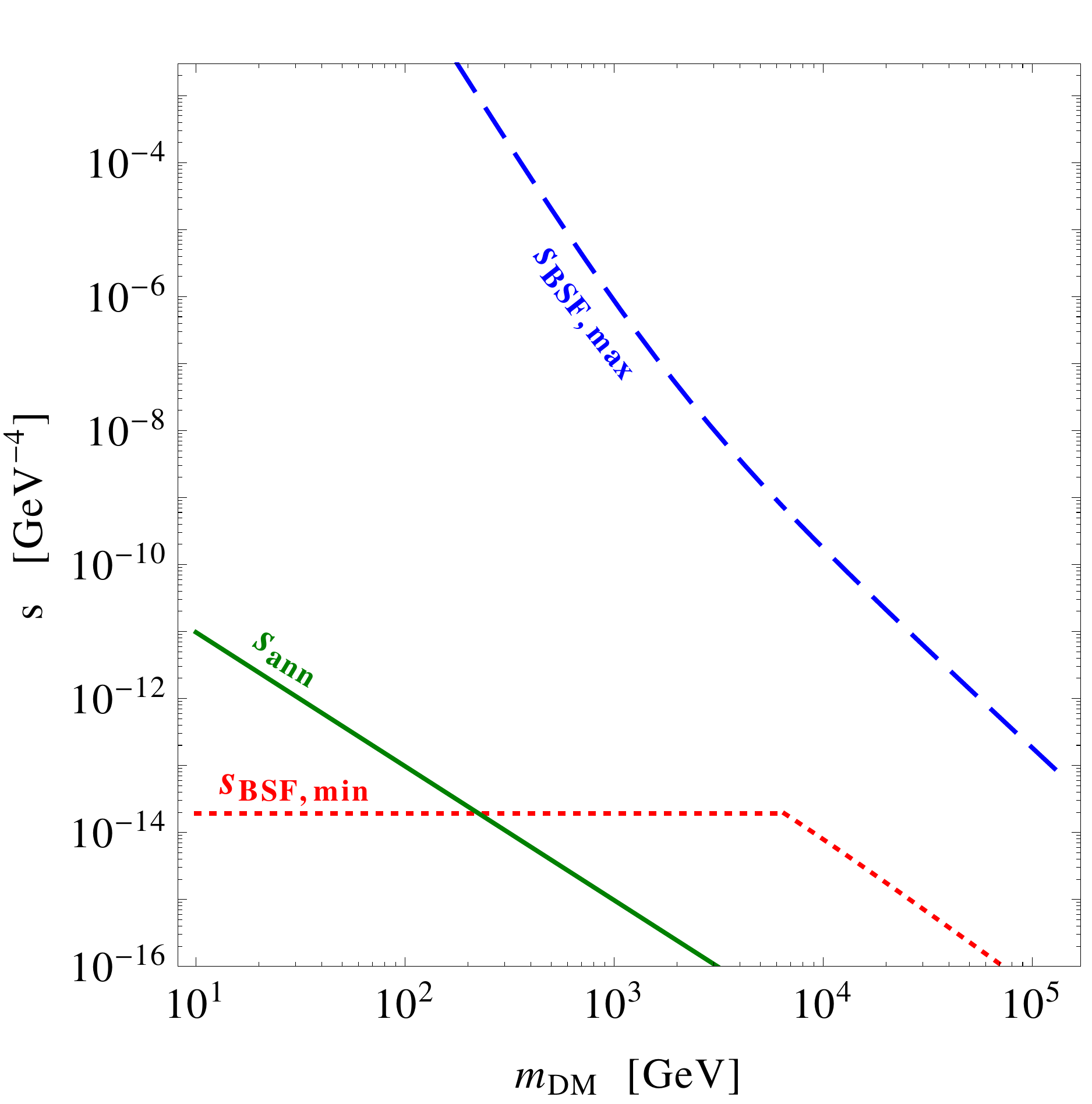}
\caption{Same as in Fig.~\ref{fig:comp_alpha}, for the signal strength as a function of the DM mass. For atomic DM, $m_{_{\rm DM}}$ is the dark Hydrogen mass.}
\label{fig:comp_mDM}
\end{figure}

The upper bound on $s\BSF$ corresponding to the partially ionized branch of Eq.~\eqref{eq:sBSF}, can be re-expressed in terms of $\mHD$, using Eq.~\eqref{eq:alpha ann} and $\mpD\geqslant \mHD/2$,
\begin{equation}
s_{\rm BSF, max} (\mHD) \approx 
\frac{3 \cdot 10^{-19}~\GeV^{-4} \ \xi\DR^2}{\aDsym^5 (\mHD/2) \, \vrel} \, . 
\label{eq:sBSFmax mH} 
\end{equation}
Moreover, using the unitarity limit of Eq.~\eqref{eq:alpha uni}, and also the consistency condition \eqref{eq:consistency}, we obtain a lower limit on $s\BSF$ as a function of $\mHD$,
\begin{multline}
s_{\rm BSF, min} (\mHD) \approx 
\\
\min \left[
\frac{2^9 \pi^2 \, \aDsym^3(\mHD/2)}{3 e^4 \, (\mHD^4/4) \ \vrel}, 
\frac{3\cdot 10^{-19}~\GeV^{-4} \xi\DR^2}{\aDuni^5 \, \vrel}
\right] .
\label{eq:sBSFmin mH}
\end{multline}
We sketch Eqs.~\eqref{eq:sBSFmax mH} and \eqref{eq:sBSFmin mH} in Fig.~\ref{fig:comp_mDM}.

\subsubsection{Signal range at fixed value of the dissipated energy}

The detectability of signals from DM related processes depends of course on the energy of the relativistic products of these processes. In contrast to the more familiar case of DM annihilation, the energy dissipated during dark atom formation is only a fraction of the dark particle mass, $\omega_\gamma \simeq \Delta \ll \mHD$. We shall thus now express the minimum and maximum signal strengths expected from dark atom formation in terms of the binding energy, $\Delta$.

\begin{figure}
\includegraphics[width=0.9\linewidth]{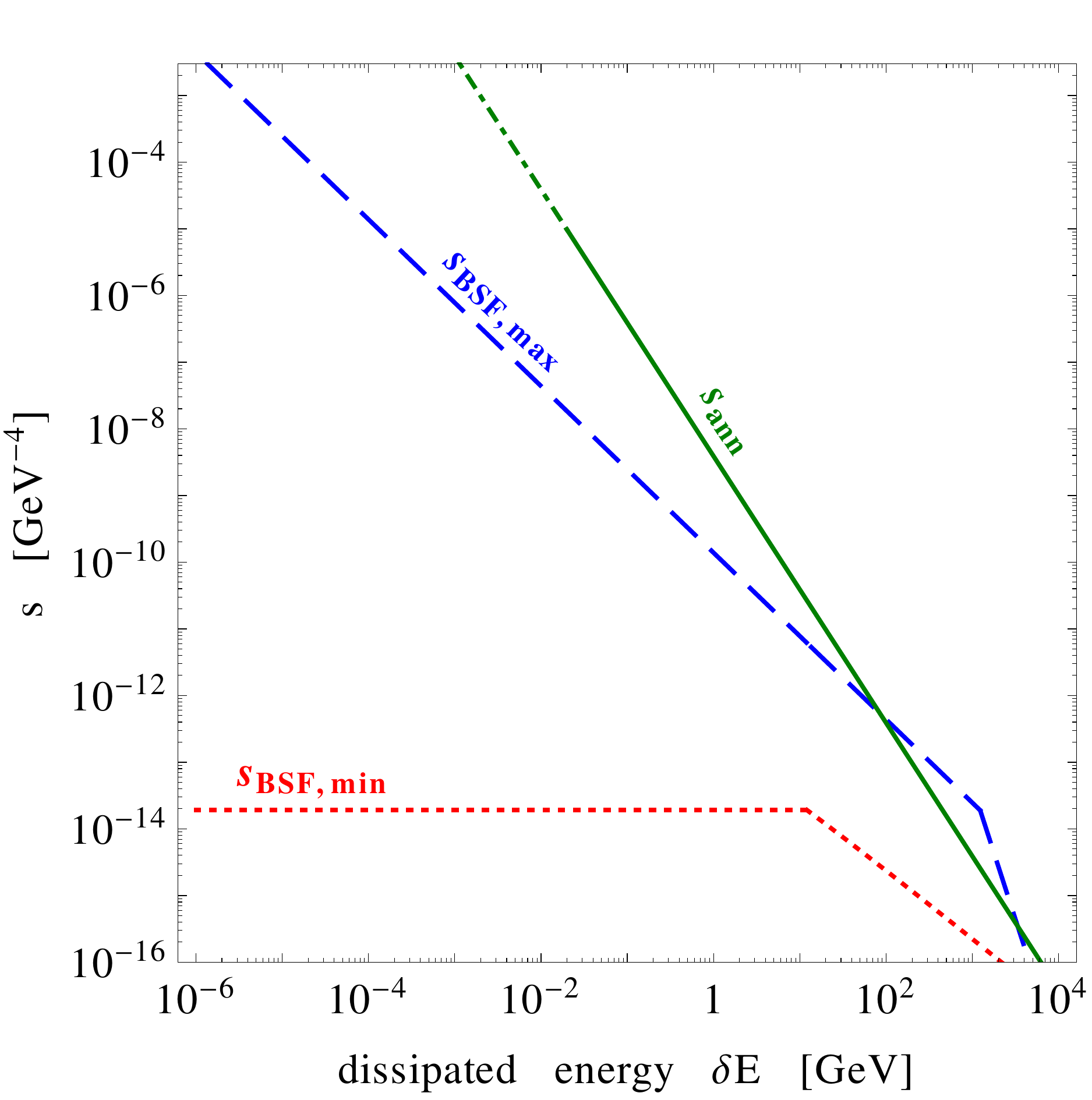}
\caption{Same as in Fig.~\ref{fig:comp_alpha}, for the signal strength as a function of the energy dissipated into radiation. For bound state formation, the dissipated energy is approximately equal to the binding energy $\delta E \simeq \Delta$, while for DM annihilation $\delta E \simeq 2 m_{_{\rm DM}}$. Note that thermal relic DM with mass $m_{_{\rm DM}} \lesssim$~few~MeV, necessitates $\xi < 1$, otherwise it would become non-relativistic and  freeze-out around or after BBN (dot-dashed green line).
}
\label{fig:comp_energy}
\end{figure}

It will be convenient to rewrite Eq.~\eqref{eq:sBSF approx}, exchanging $\muD$ for $\Delta$, as follows
\begin{equation}
s\BSF \approx \min \left[
\frac{2^7 \pi^2 \, \aD^7}{3e^4 \, \mHD^2 \Delta^2 \vrel},
\frac{3 \cdot 10^{-19}~\GeV^{-4} \: \xi\DR^2}{\aD^5 \vrel} 
\right] .
\label{eq:sBSF approx 2}
\end{equation}
For a fixed value of $\mHD\Delta$, the maximal signal is
\begin{equation}
s_{\rm BSF, max} \approx 3.6 \times 10^{-11} \, \GeV^{-4} \ \frac{\xi\DR^{7/6}}{\vrel}
\left(\frac{\GeV^2}{\mHD \Delta}\right)^{5/6} \, ,
\label{eq:sBSFmax mH Delta} 
\end{equation}
and occurs if 
\begin{equation}
\xi\DR \left(\frac{\mHD \Delta }{\GeV^2}\right) \approx 5 \times 10^9 \: \aD^6 \, .
\label{eq:smax condition}
\end{equation} 
From Eq.~\eqref{eq:masses range}, $\aD^2 \gtrsim 8\Delta/\mHD$, which together with Eq.~\eqref{eq:smax condition} implies $\xi\DR^{1/4}(\mHD/\GeV) \gtrsim 40 (\Delta/\MeV)^{1/2}$ along the line of maximal signal. Equation~\eqref{eq:sBSFmax mH Delta} now becomes
\begin{equation}
s_{\rm BSF, max} \approx 
5.3 \times 10^{-10}~\GeV^{-4} \ \frac{\xi\DR^{11/8}}{\vrel} \left(\frac{\MeV}{\Delta}\right)^{5/4}
\, .
\label{eq:sBSFmax Delta1} 
\end{equation}
This is valid provided that $8\Delta / \mHD \lesssim \aD^2 \lesssim \aDuni^2$. For $\mHD \gtrsim 8 \Delta/\aDuni^2$, the fully ionized branch of Eq.~\eqref{eq:sBSF approx 2} can give a stronger bound. 
Collecting everything, we find
\begin{multline}
s_{\rm BSF, max} (\Delta)\approx \min\left[
\frac{2\pi^2 \, \aDuni^{11}}{3e^4 \, \vrel \, \Delta^4} , 
\right. \\  \left.
5.3 \times 10^{-10}~\GeV^{-4} \ \frac{\xi\DR^{11/8}}{\vrel} \left(\frac{\MeV}{\Delta}\right)^{5/4}
\right] \, .
\label{eq:sBSFmax Delta} 
\end{multline}
We obtain lower bounds on $s\BSF$ by considering the fully and partially ionized branches of Eq.~\eqref{eq:sBSF approx 2} separately, and taking into account that $\mHD < 2\mpD < 2\mpDsym (\aD)$. We obtain
\begin{multline}
s_{\rm BSF, min} (\Delta) \approx \min \left\{
\frac{3 \cdot 10^{-19}~\GeV^{-4} \xi\DR^2}{\aDuni^5 \, \vrel}, 
\right. \\ \left.
\frac{2^5 \pi^2}{3 e^4 \, \Delta^2 \ \vrel} 
\left[\frac{\aD^7}{\mpDsym^2(\aD)}\right]_{\aD \to \aDsymT(\Delta)}
\right\} \, .
\label{eq:sBSFmin Delta} 
\end{multline}
We sketch Eqs.~\eqref{eq:sBSFmax Delta} and \eqref{eq:sBSFmin Delta} in Fig.~\ref{fig:comp_energy}.

\medskip

We note that Eqs.~\eqref{eq:sBSFmax alpha} -- \eqref{eq:sBSFmin mH}, \eqref{eq:sBSFmax Delta} and \eqref{eq:sBSFmin Delta} assume $\aD > \vrel$; for $\aD < \vrel$, similar considerations can be carried out, albeit the signals from bound-state formation weaken significantly. 
In the above, we used the same symbols ($s_{\rm BSF, min}$ and $s_{\rm BSF, max}$) for different functions to ease the notation.

\subsection{Comparison to annihilating dark matter \label{sec:BSF-ann comparison}}

We now compare the expected signal from  dark atom formation, with the expected signal from annihilation of symmetric thermal relic DM;\footnote{Asymmetric DM may also result in suppressed but detectable annihilation signals due to the subdominant population of antiparticles left over from the early universe~\cite{Hardy:2014dea,Bell:2014xta}. Moreover, annihilation signals are predicted in scenarios where DM possessed a particle-antiparticle asymmetry at early times that was subsequently erased due to oscillations~\cite{Cirelli:2011ac}. We do not consider these possibilities here.}
the latter includes the case of WIMP DM, however symmetric thermal relic DM may also reside in a hidden sector~\cite{Pospelov:2007mp}. The rate of annihilations per volume is $d\Gamma_{\rm ann}/dV = \rho\DM^2 \, s_{\rm ann}$, where for non-self-conjugate DM
\begin{equation}
s_{\rm ann} = \dfrac{(\sigma \vrel)_{\rm ann}}{4m_\chi^2} \: ,
\label{eq:sann}
\end{equation}
For $s$-wave annihilation, the cross-section which yields the observed DM density is~\cite{Steigman:2012nb}
\begin{equation}
(\sigma \vrel )_\mathrm{ann} \simeq \xi\FO \times 4.4 \times 10^{-26} \; \mathrm{cm}^3 \slash \mathrm{s} \: .
\label{eq:canonical}
\end{equation}
Here we consider only DM annihilating via a short-range interaction; for DM interacting via a long-range force, the annihilation cross-section is enhanced at low velocities by the Sommerfeld effect, which may affect both the freeze-out~\cite{Feng:2009hw,vonHarling:2014kha} and the annihilation signals today~\cite{ArkaniHamed:2008qn,Pospelov:2008jd,MarchRussell:2008tu,Shepherd:2009sa}. 
We compare $s_{\rm ann}$ to $s_\mathrm{BSF,min}$ and $s_\mathrm{BSF,max}$ in Figs.~\ref{fig:comp_alpha} -- \ref{fig:comp_energy}.

In Fig.~\ref{fig:comp_alpha}, we present $s_\mathrm{BSF,min}$, $s_\mathrm{BSF,max}$ and $s_{\rm ann}$ vs. the DM coupling which is responsible for the process that produces the expected signals. For atomic DM, this is the dark fine structure constant $\aD$. To facilitate the comparison, for the symmetric DM case, we shall consider Dirac fermions coupled to a massive dark photon that is lighter than the DM particles albeit sufficiently heavy such that the Sommerfeld effect is not relevant. In this case, $(\sigma \vrel)_{\rm ann} \approx \pi \aD^2/ m\DM^2$. Then, the required value of the annihilation cross-section quoted in Eq.~\eqref{eq:canonical} implies a relation between $\aD$ and $m\DM$, which we use in sketching the $s_{\rm ann}$ curve in Fig.~\ref{fig:comp_alpha}. We see that, for the same coupling $\aD$, the signal expected from the formation of dark atoms exceeds significantly the signal expected from the annihilation of symmetric thermal-relic DM. This is partly due to the Sommerfeld enhancement of bound state formation, which renders it very efficient at the low relative velocities inside DM haloes.

Similarly, when compared at equal values of the DM mass, dark atom formation appears to be more efficient than the annihilation of symmetric DM in much of the parameter space, as seen in Fig.~\ref{fig:comp_mDM}. This is in part because for symmetric thermal-relic DM, the DM coupling strength is fixed by Eq.~\eqref{eq:canonical}; in contrast, in the asymmetric DM scenario, stronger annihilation does not diminish the DM abundance and there is no upper bound (but rather only a lower one) on the DM coupling from this consideration. Conversely, a given value of $\aD$ may correspond to much smaller DM mass in the case of asymmetric DM, than in the case of symmetric DM; smaller mass implies, in turn, larger interaction cross-section, as seen in Fig.~\ref{fig:comp_alpha}. Moreover, $\sigma\BSF \propto \muD^{-2}$ while $\sigma_{\rm ann} \propto m\DM^{-2}$, and in general, $\muD$ may be much smaller than $\mHD$.

In Fig.~\ref{fig:comp_energy}, we compare the signal strengths at equal values of the dissipated energy. For atomic DM, $\delta E \simeq \Delta \ll \mHD$, while for annihilating DM, $\delta E = 2m\DM$. At small $\epsilon$, the annihilation of symmetric thermal relic DM is expected to give stronger signals than dark atom formation, despite the latter being a more efficient process. This is because dark atom formation is suppressed by the small number density of the dark particles, due to the much larger DM mass (for the same value of $\delta E$). However, annihilating species cannot be cold thermal relics if their mass is significantly below $\sim 10$~MeV, unless $\xi \ll 1$. It is thus unlikely that the annihilation of thermal relic DM can explain sharp features in the low-energy part of the $\gamma$-ray spectrum, such as the observed 511~keV and 3.55~keV lines. 
On the other hand, thermal relic DM can potentially account for such features if it is asymmetric and possesses internal structure which gives rise to level transitions~\cite{Frandsen:2014lfa,Cline:2014eaa,Boddy:2014qxa}, such as formation of bound states. We discuss in particular, the possibility of the 511~keV line being produced due to the radiative formation of dark atoms, in the next section. 
At larger values of the dissipated energy $\delta E$ (which also implies larger DM mass and larger coupling), the signals from dark atom formation can be potentially comparable (or stronger) to those expected from symmetric DM annihilation as seen in Fig.~\ref{fig:comp_energy}.

\medskip

The above discussion does not capture, of course, the entire complexity of the indirect detection of DM.
The detectability of the radiation produced by DM-related processes depends not only on the rate and the total energy dissipated, but also on the energy spectrum and the nature of the end products. 
In the radiative formation of dark atoms shown in \eqref{eq:BSF}, after averaging over the polarizations of the massive photon, the dark photon decay products are equally likely to be emitted in any direction with a flat energy spectrum, extending to energies $E \sim \Delta$. In contrast, the direct products of DM annihilation have a sharply peaked spectrum at energy $E \sim m\DM$. However, most direct WIMP annihilation products, e.g.~$b\bar b$, are followed by cascade decays which result in an extended final spectrum at lower energies.  In addition, propagation effects can substantially modify the spectrum shape for both WIMP annihilation and dark atom formation.

\section{The 511 keV line from dark atom formation in the galactic center \label{sec:511 keV}}

In this section we focus on charged decay products of the dark photon, which are more easily detectable than neutrinos. (The $\gammaD \to 3\gamma$ decay is extremely suppressed, as seen from Eq.~\eqref{eq:gammaD to 3gamma}.)  We saw above that, when the dark photon decay into charged particles is kinematically allowed, at $\Delta \gtrsim \MeV$, the largest signal strength occurs near $\aD \sim 0.015$ and $\mHD\muD \sim 400~\GeV^2$.  The consistency condition \eqref{eq:consistency}, then implies $\muD \lesssim 10~\GeV$ and correspondingly, $\mHD \gtrsim 40~\GeV$.  (Of course, allowing for somewhat lower signals broadens the parameter range.) 
The requirement $1.022~\MeV < M_\gamma < \Delta$ ensures that close to the maximal signal region, appreciable signals are produced only in the $e^+e^-$ channel. We shall now explore the possibility that in this region -- of near-maximal signal and low positron injection energy $E_{\rm inj} \lesssim \Delta/2 \sim~\MeV$ -- the formation of dark atoms may explain the 511~keV line observed in the center of the Milky Way~\cite{Knodlseder:2005yq}.

\subsection{The 511 keV line \label{sec:511 keV - general}}

The 511~keV photon flux from the bulge of the Milky Way is observed to be
$(1.05 \pm 0.06) \times 10^{-3}~{\rm photons \ cm^{-2} \, s^{-1}}$, with a spatial extension of $\sim 8^\circ$ (FWHM)~\cite{Knodlseder:2005yq}. This corresponds to an annihilation rate of non-relativistic positrons of
\begin{equation}
\Gamma_{e^+e^-, \rm obs} \approx 1.5 \times 10^{43} \; \mathrm{s}^{-1} \, ,
\label{eq:511 flux obs}
\end{equation}
within radius $r_\mathrm{max} \approx (8^\circ \slash 360^\circ) \pi R_{\rm sc} \simeq 0.6~\mathrm{kpc}$ from the galactic center, where $R_{\rm sc} \simeq 8.5~\mathrm{kpc}$ is the radius of the solar circle.~\footnote{In deducing the positron annihilation rate of Eq.~\eqref{eq:511 flux obs} from the observed photon flux, the positronium fraction in the galaxy was taken into account. For details see~\cite{Knodlseder:2005yq,Beacom:2005qv}.} 

Positrons injected in the Galaxy at some higher energy, propagate in the interstellar medium, lose energy, and annihilate with electrons. The majority of positrons survive until they become non-relativistic and annihilate at rest to produce the 511 keV line. However, a portion of the positrons annihilate while still relativistic. The survival probability is $P_{\rm surv} \simeq 0.95$~\cite{Beacom:2005qv,Bell:2010fk}, with a very mild energy dependence which we shall ignore.

The in-flight positron annihilations can produce a significant flux of $\gamma$-rays; the observed $\gamma$-ray continuum then constrains the injection energy of the positrons responsible for the observed 511~keV line. Monoenergetic positrons should be injected at energies $E_{\rm inj} \lesssim 3~\MeV$~\cite{Beacom:2005qv}, although this bound could be relaxed if uncertainty greater than 30\% in the diffuse $\gamma$-ray flux is incorporated.\footnote{A weaker constraint, $E_{\rm inj} \lesssim 20~\MeV$, arises from the internal bremsstrahlung photons associated with any process which produces charged particles~\cite{Beacom:2004pe}.} 
In addition, a broader positron injection distribution, such as that produced by the dark photon decay, could extend to somewhat higher energies.

\subsection{Positrons from dark atom formation}

While astrophysical explanations for the production of the galactic positrons have been proposed~\cite{Casse:2003fh,Bertone:2004ek,Parizot:2004ph,Prantzos:2010wi},  the origin of the 511 keV line remains a mystery.
Dark matter related explanations have also been put forward (see e.g.~\cite{Finkbeiner:2007kk}). Here we show that the decay of dark photons emitted in the formation of dark atoms in the center of the Milky Way, can account for the observed positron flux. 

The rate at which dark atom formation can contribute to the non-relativistic positron annihilation in the center of the galaxy is
\begin{equation}
\Gamma_{e^+, \rm BSF} \approx  s\BSF \times I \times P_{\rm surv} \, ,
\end{equation}
where 
\begin{equation}
I = \int dV \, \rho\DM^2 \, ,
\end{equation}
with the integration extending to radius $r_{\rm max}$ from the center of the galaxy. 
We parametrize the DM density profile by
\begin{equation}
\rho\DM (r) = \dfrac{\rho_0}{\left(\dfrac{r}{r_s}\right)^\gamma \left[ 1 + \left(\dfrac{r}{r_s}\right)^\alpha \right]^{(\beta - \gamma) \slash \alpha}},
\label{eq:DM_profile}
\end{equation}
where $r_s =$~20~kpc and $\rho_0$ is determined such that $\rho(R_{\rm sc}) = 0.4~\GeV \slash \mathrm{cm}^3$.  The Navarro-Frenk-White profile corresponds to $\alpha = 1$, $\beta= 3$, and $\gamma = 1$.  However, the slope may be steeper in the galactic center due to the supermassive black hole~\cite{Gondolo:1999ef} and/or baryonic matter~\cite{Blumenthal:1985qy,Gnedin:2003rj,Gustafsson:2006gr}, as has been suggested by numerical simulations~\cite{Gnedin:2011uj}.  We fix $\alpha = 1$ and $\beta = 3$, but consider $1< \gamma <1.4$. For this range of $\gamma$, $I \simeq (0.1 - 3.3)\times 10^{68}~\GeV^2~\cm^{-3}$. Then, the signal strength required to produce the rate of Eq.~\eqref{eq:511 flux obs} is
\begin{equation}
s_{_{\rm BSF, 511}} \approx  1.4 \times 10^{-8}~\GeV^{-4} \ \left(\frac{10^{68}~\GeV^2\,\cm^{-3}}{I}\right) \, .
\label{eq:511 sBSF required}
\end{equation}

\begin{figure}[t!]
\includegraphics[scale=.6]{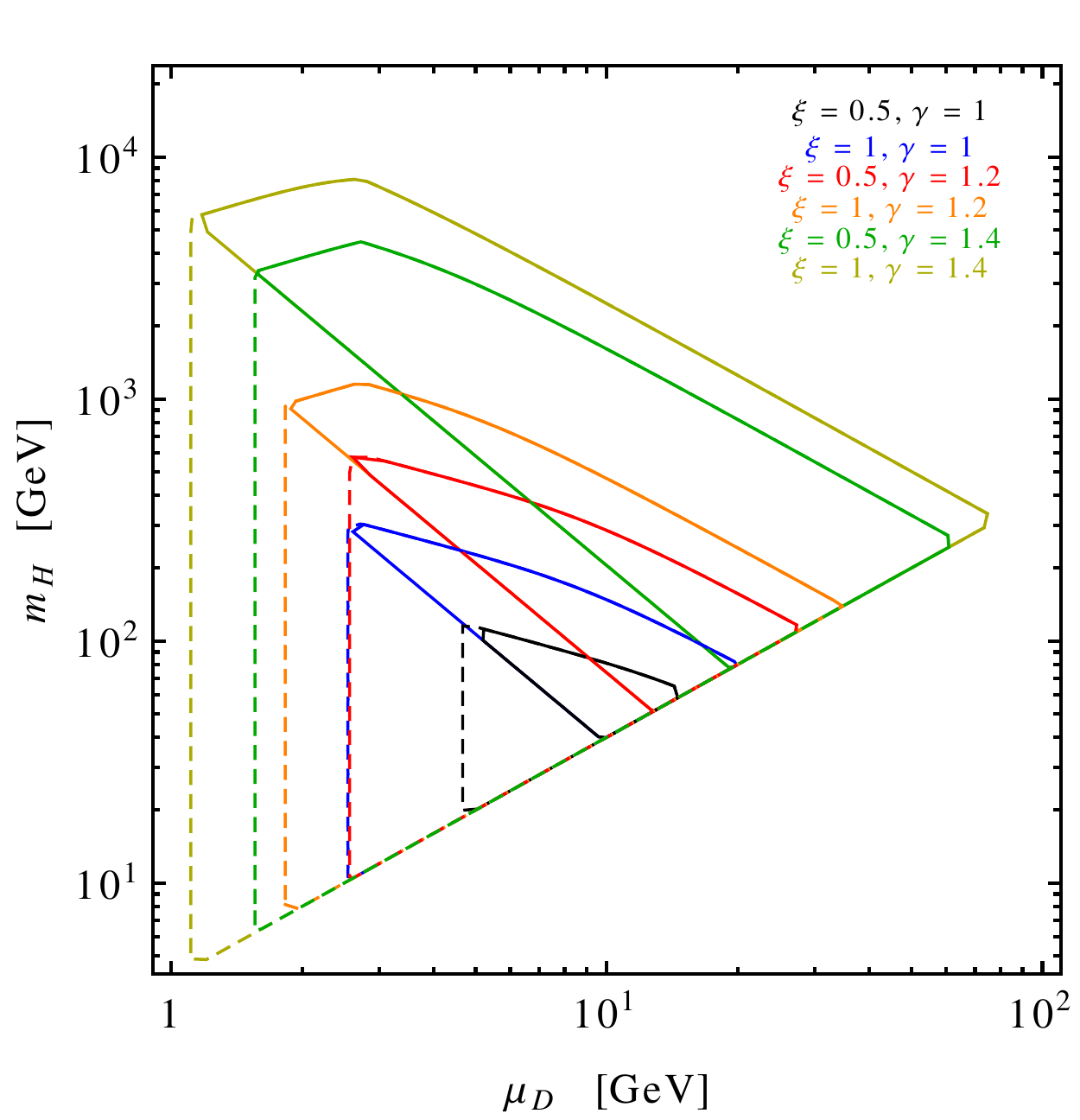}
\includegraphics[scale=.6]{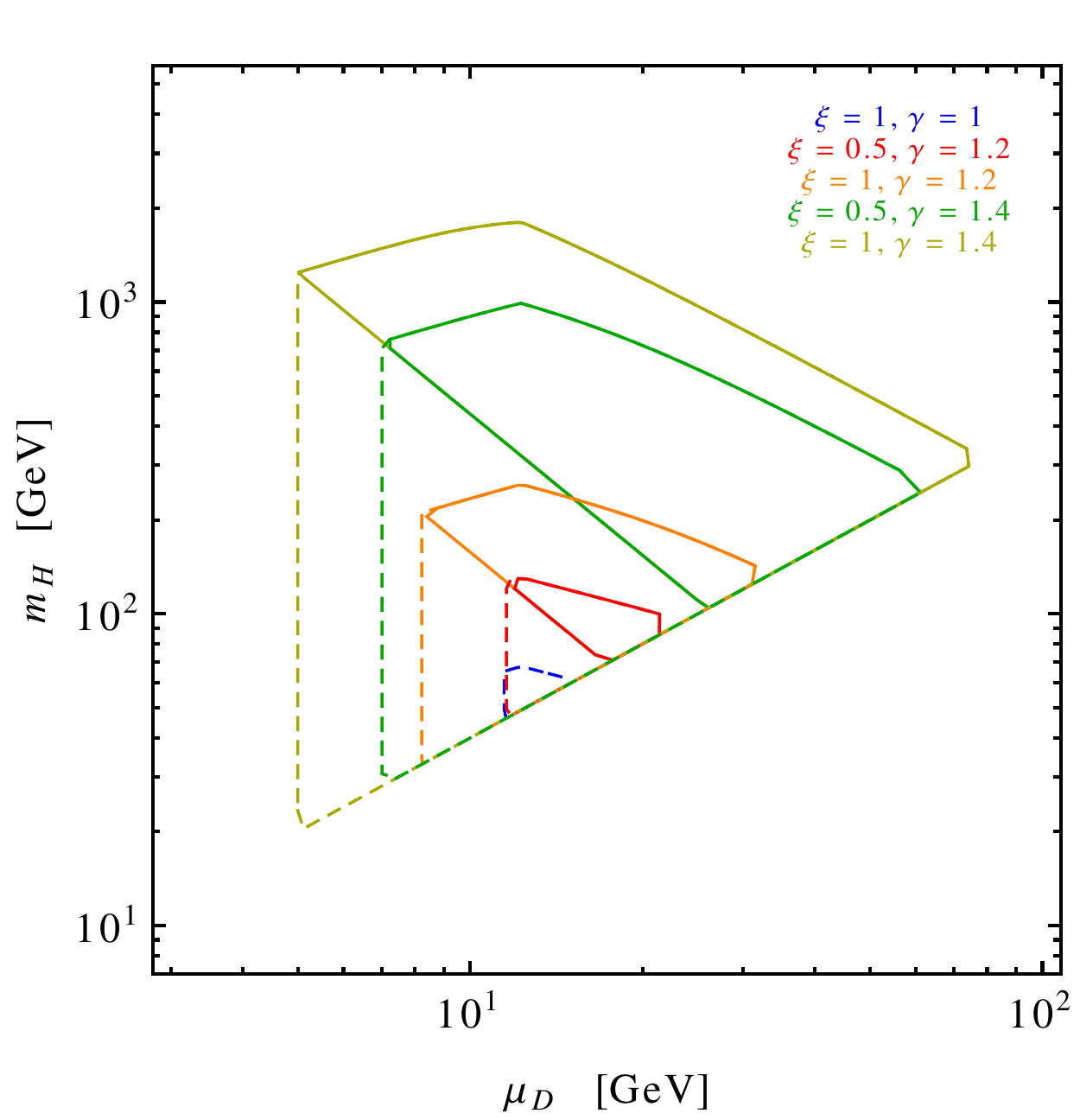}
\caption{Parameter space which can produce the 511 keV line, for an appropriate choice of $\alpha_{_D}$ which depends on $m_{_{\rm H}}$, $\mu_{_D}$. The dark photon mass is $M_\gamma = 2 \; \mathrm{MeV}$ (top) and $M_\gamma = 9~\mathrm{MeV}$ (bottom).  The solid lines correspond to fully ionized DM, while the dashed lines correspond to partially ionized DM. Note that the dashed lines encompass all of the region enclosed by the solid lines, for the same $\xi$ and $\gamma$, albeit the values of $\alpha_{_D}$ are different in the two cases. (As seen from Fig.~\ref{fig:signal_1}, the maximum value of $m_{_{\rm H}}\mu_{_D}$ which can produce a given signal $s_{_{\rm BSF}}$, is the same for fully and partially ionized DM.)
More contracted halo profiles (larger $\gamma$) and larger dark-to-ordinary temperature ratios $\xi$ in the early universe, lead to a stronger signal and hence a larger parameter space.}
\label{fig:param_space}
\end{figure}

In Fig.~\ref{fig:param_space} we show the boundaries of the parameter space that can account for the observed excess, for various choices of dark photon mass $\Mg$, and the ratio of the dark to visible temperatures $\xi$.  For each set of values ($\muD, \, \mHD$), we look for the value of $\aD$ that yields the observed rate.  Because the signal strength produced by dark atom formation is bounded, as discussed in Sec.~\ref{sec:signal boundaries}, there is not always a value of $\aD$ that can produce the observed rate.  In Fig.~\ref{fig:param_space}, we also require that the consistency relation~\eqref{eq:consistency} is satisfied, and that dark atoms can form radiatively, i.e.~$M_\gamma < \Delta$. Note that in this region of parameter space, the phase-space suppression factor is not negligible, and we do incorporate it in our calculations. These plots do not include astrophysical and cosmological constraints, which will be discussed below.

Since dark atom formation is more rapid in more contracted profiles (large $\gamma$), the parameter space which can account for the 511~keV flux is correspondingly larger.  Moreover, a larger temperature ratio $T_D/T_V$, increases the residual DM ionization fraction and strengthens the expected signal, thus proving larger parameter space for explaining the observed flux. The available parameter space is also dependent on the mass of the dark photon; if $\Mg$ is close to the binding energy, the phase space suppression of the signal limits the parameter space which can account for the line.

For a fixed reduced mass $\muD$ and bound state mass $\mHD$, there are typically two couplings $\aD$ which can produce the observed 511 keV flux; one corresponds to fully ionized DM today, while the other corresponds to partially ionized DM today.  The couplings necessary to produce the observed flux are generally smaller in the partially ionized branch; therefore, less parameter space is eliminated by the constriant $\Delta > \Mg$, and consequently, the parameter space in which the observed 511 keV excess can be produced is generally larger for the partially ionized branch.

We must also ensure that the dark photons emitted during the formation of dark atoms in the galactic center, decay into $e^+e^-$ within $r \lesssim r_{\rm max}$. From Eq.~\eqref{eq:gammaD to ff}, the dark photon decay length is $\lambda \simeq [\epsilon^2 \alpha_{_{\rm EM}} \Mg^2 /(3\Delta)]^{-1}$, where we took into account the dark photon Lorentz boost $\Delta/\Mg$. Requiring conservatively $\lambda \lesssim r_{\rm max}/10\simeq 0.06~{\rm kpc}$, sets a very comfortable lower bound on the kinetic mixing
\begin{equation}
\epsilon \gtrsim 10^{-14} \left(\frac{2~\MeV}{\Mg}\right)
\left(\frac{10~\MeV}{\Delta}\right)^{1/2} \, .
\label{eq:epsilon min}
\end{equation}

\subsection{Astrophysical and Cosmological Constraints}

\begin{figure}
\includegraphics[scale=.5]{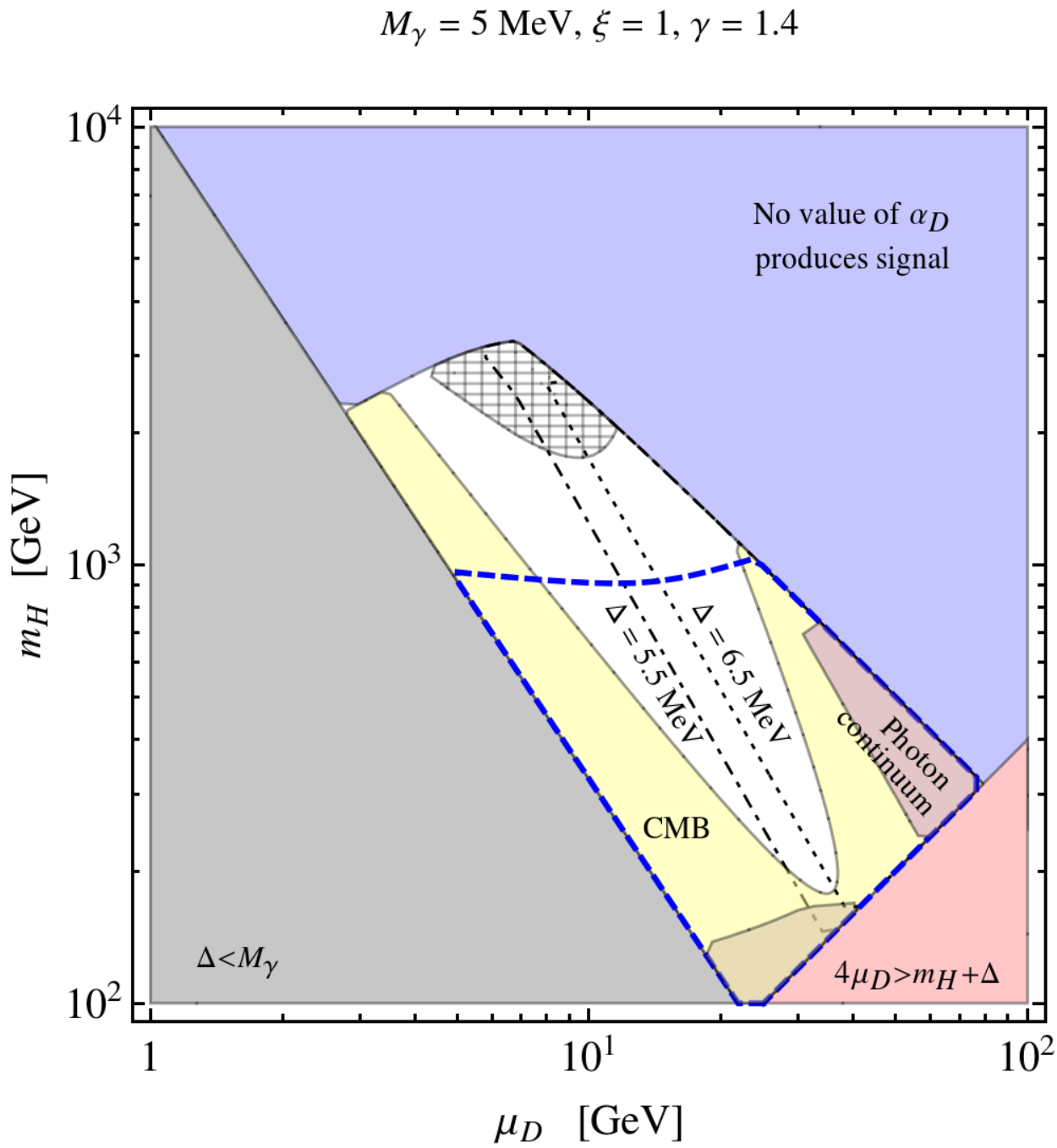}
\includegraphics[scale=.5]{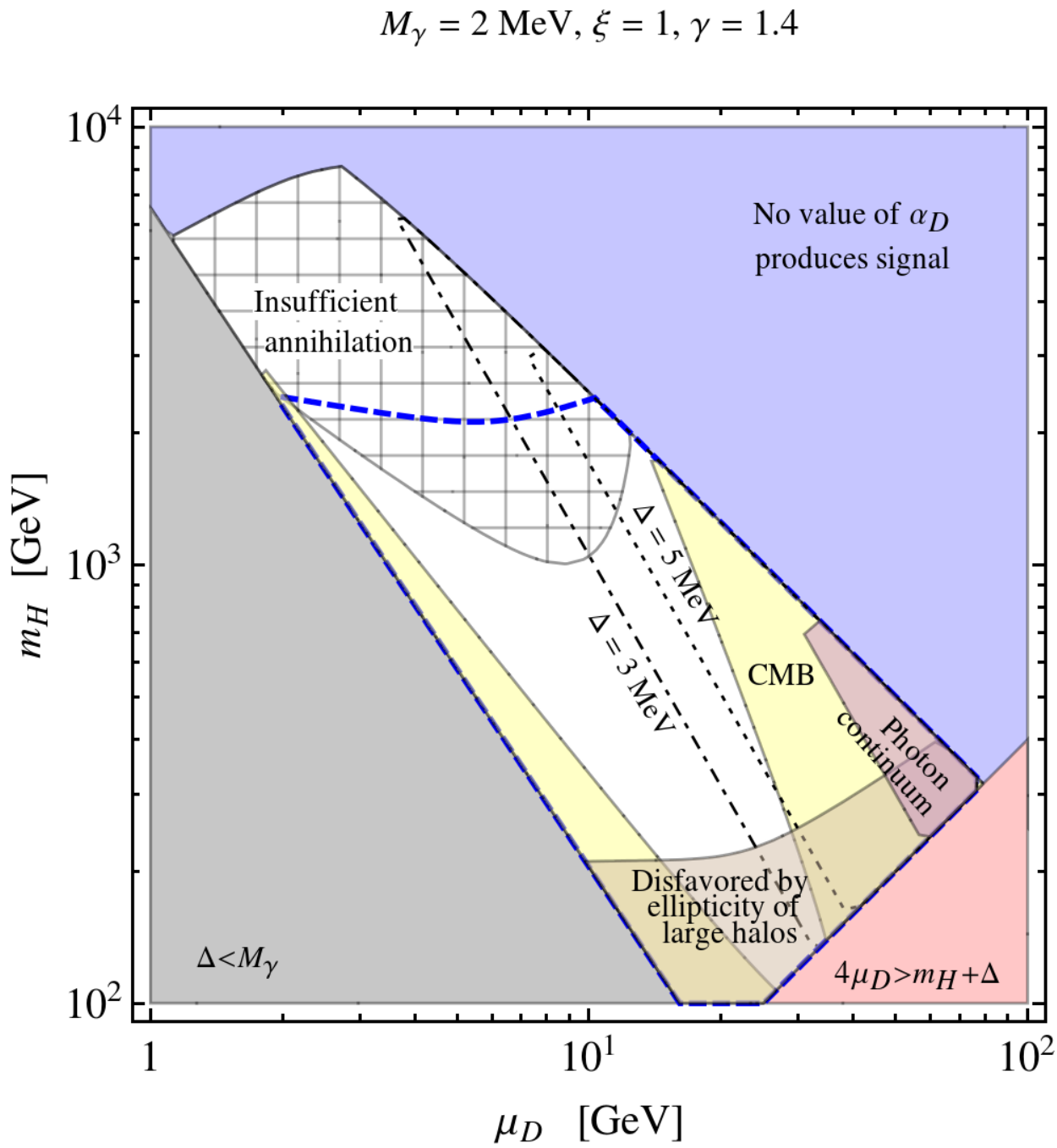}
\includegraphics[scale=.5]{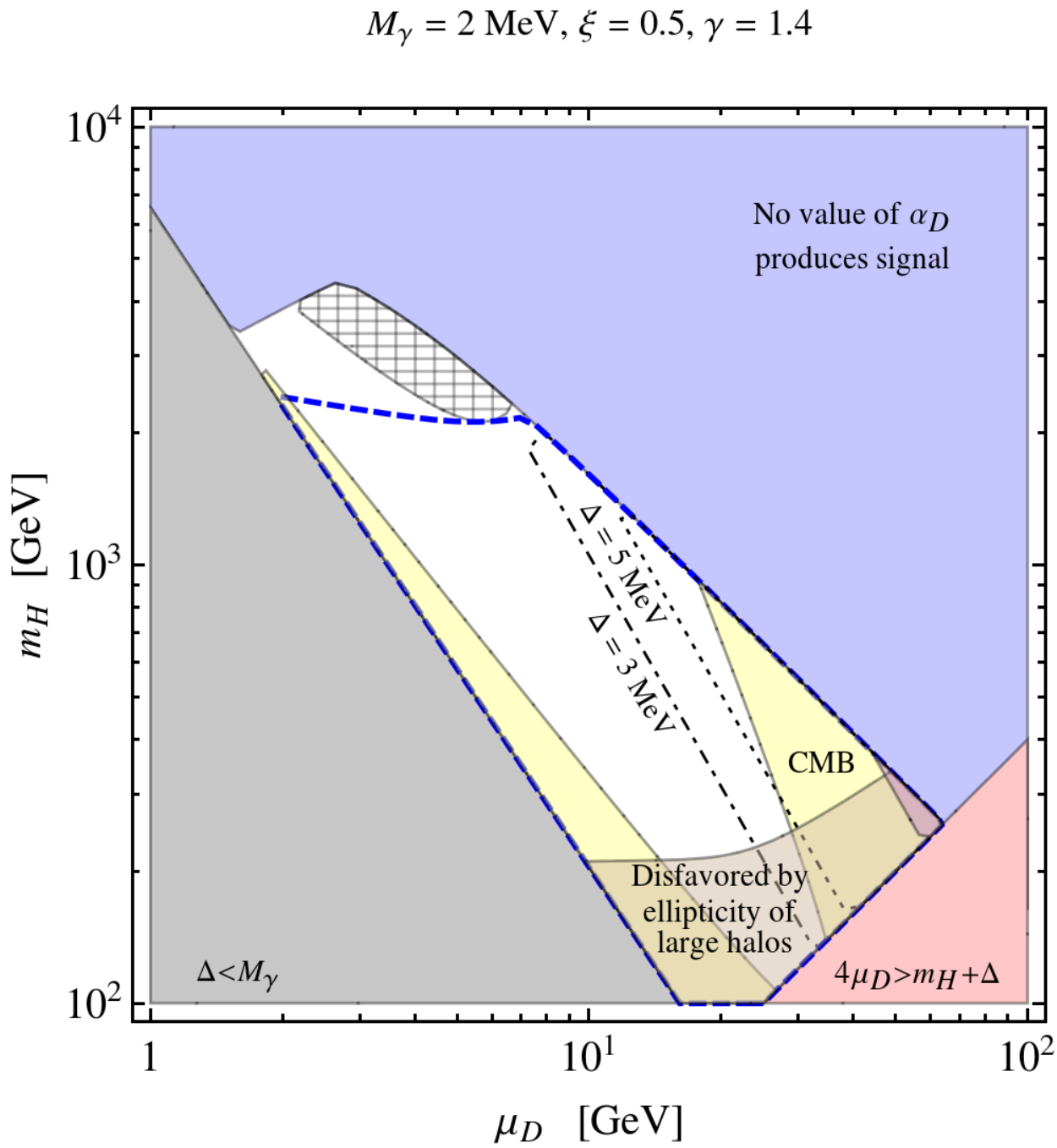}
\caption{The parameter space which can account for the 511~keV line (white region), within the regime where DM is fully ionized, $x_{_D} = 1$, and for a contracted NFW profile with $\gamma = 1.4$; the dark photon mass $M_\gamma$, and the early-universe dark-to-ordinary temperature ratio $\xi$, are denoted on the top of each plot. We have applied various constraints, as described in the text and by the corresponding labels. In addition, the blue dashed line encloses the region where DM self-scattering can affect the dynamics of dwarf-galaxy-sized haloes. Contours of constant binding energy $\Delta$ are also shown.
}
\label{fig:param_space_with_constraints}
\end{figure}
\begin{figure}
\includegraphics[scale=.5]{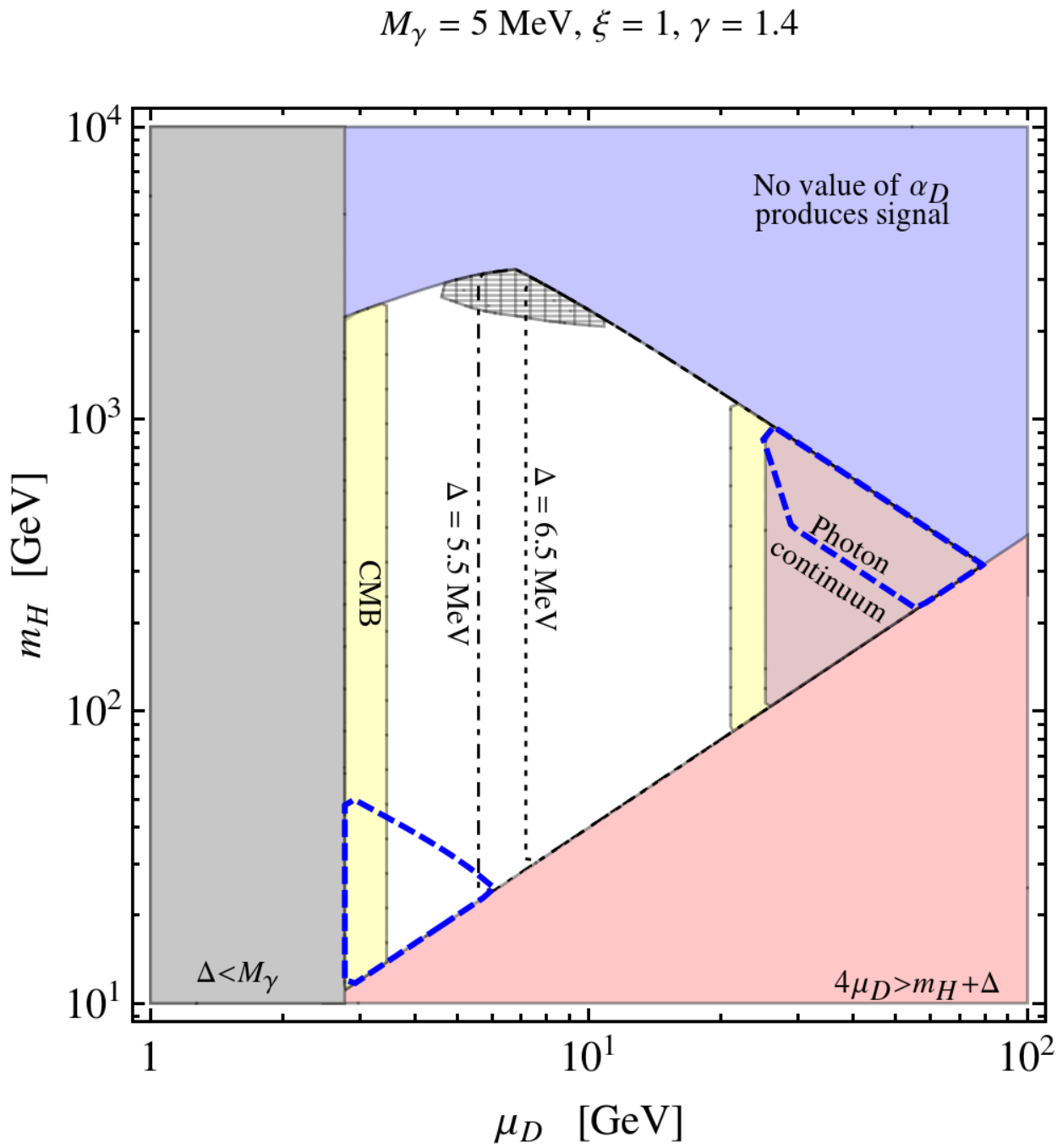}
\includegraphics[scale=.5]{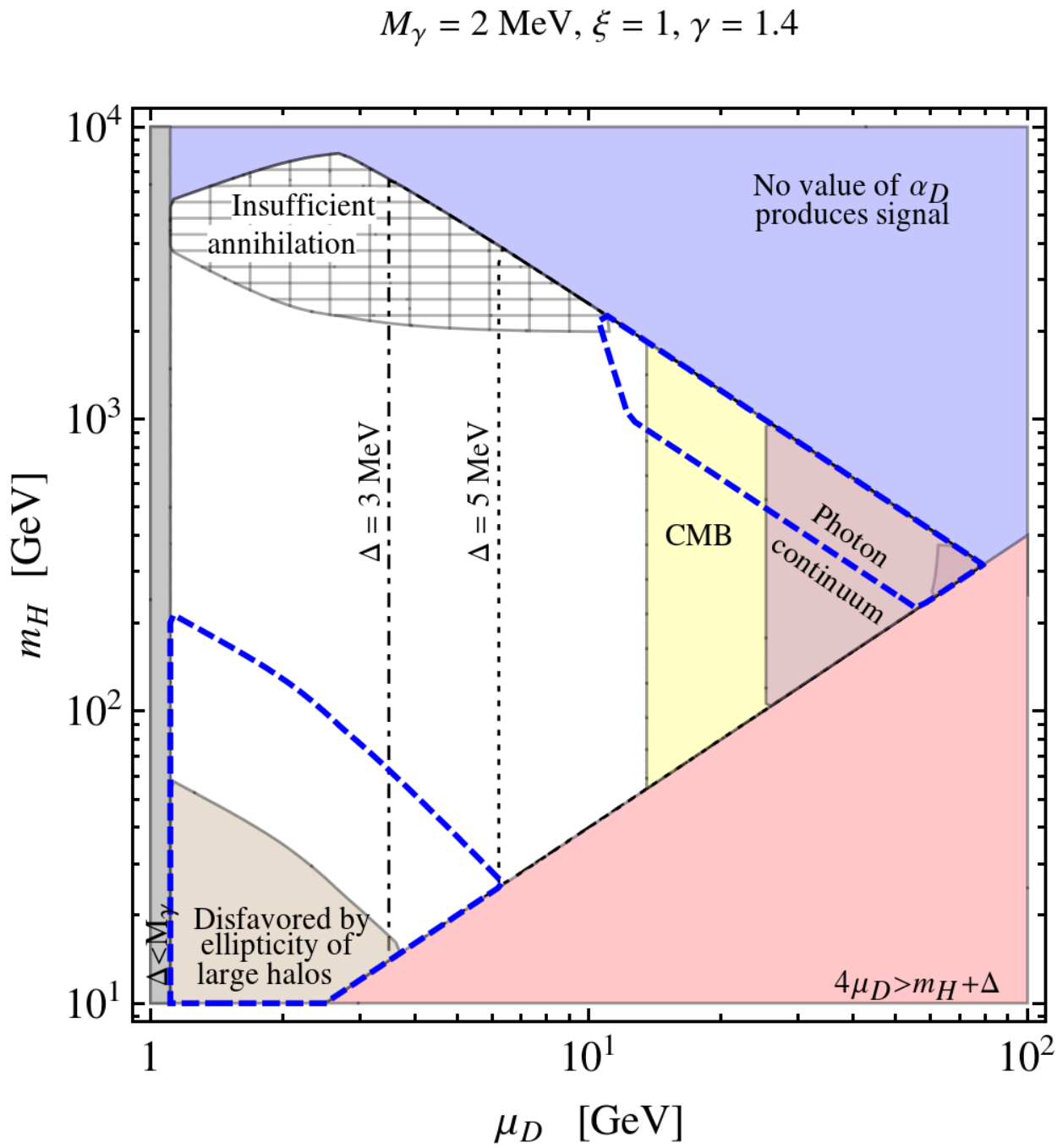}
\includegraphics[scale=.5]{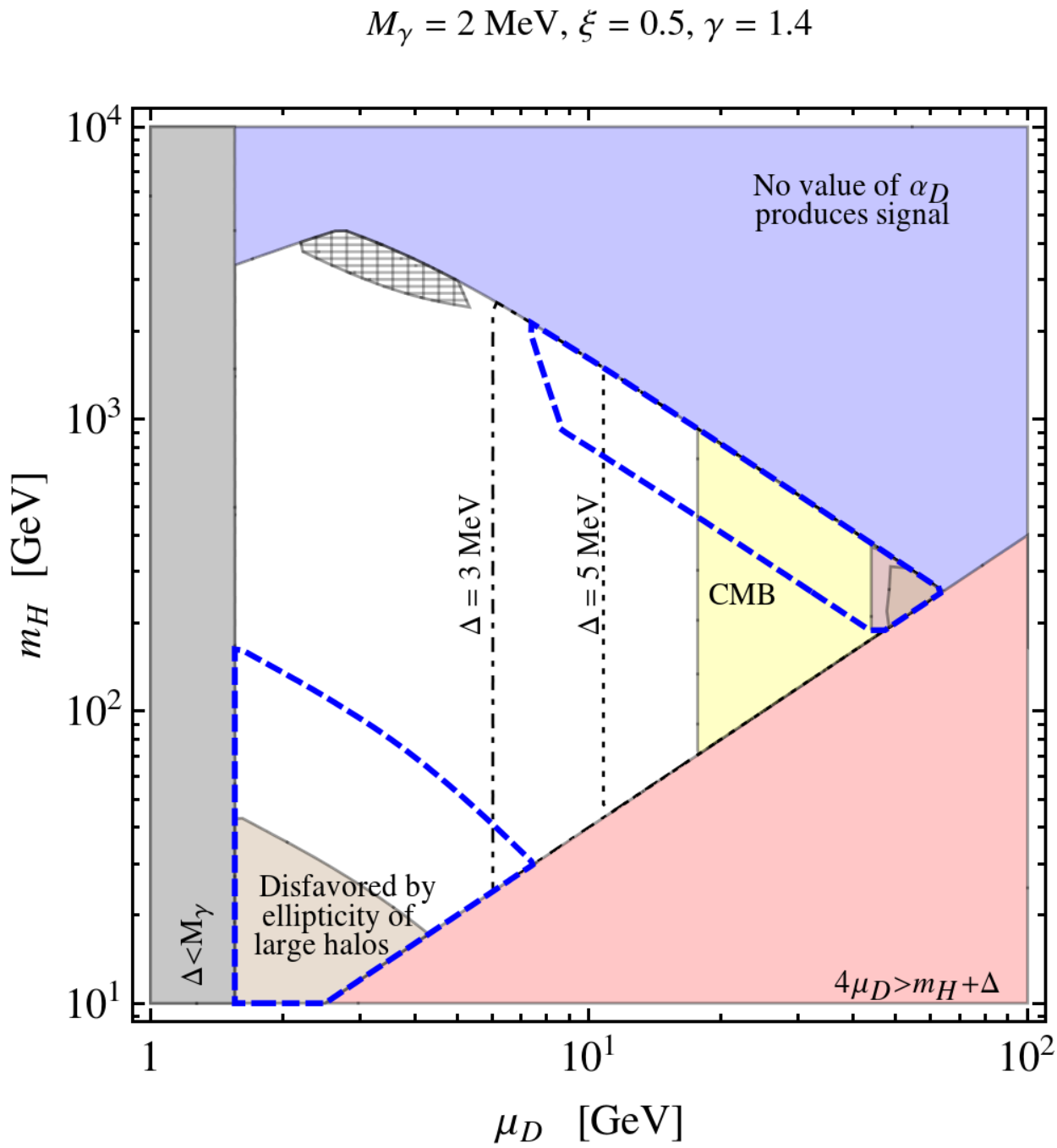}
\caption{Same as in Fig.~\ref{fig:param_space_with_constraints}, for the regime where DM is partially ionized, $x_{_D} < 1$.}
\label{fig:param_space_with_constraints_partial}
\end{figure}

\begin{figure}
\includegraphics[scale=.5]{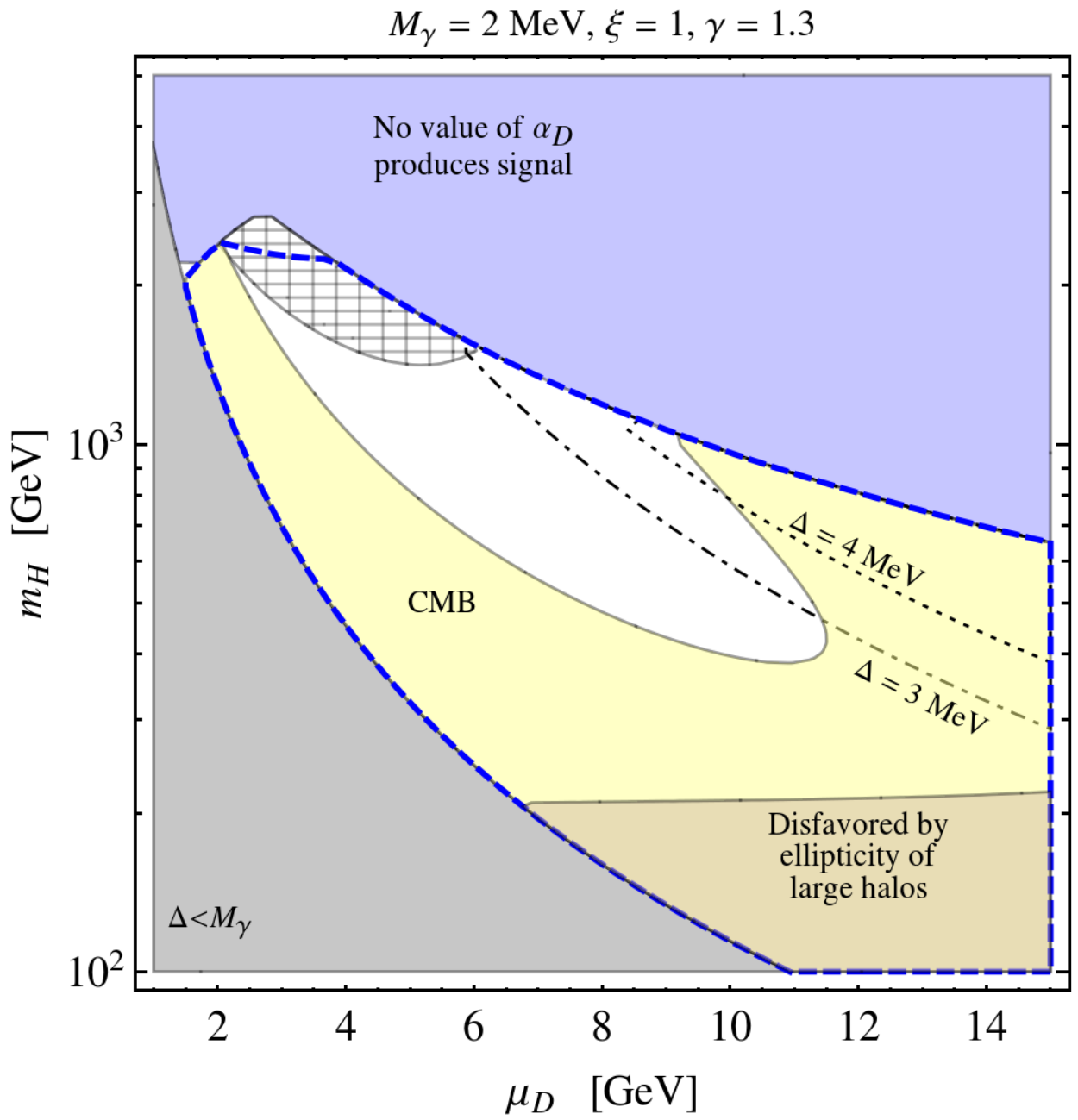}
\caption{Same as in Fig.~\ref{fig:param_space_with_constraints} (fully ionized dark matter), for a contracted NFW profile with $\gamma = 1.3$. }
\label{fig:param_space_with_constraints_2}
\end{figure}

\begin{figure}
\includegraphics[scale=.5]{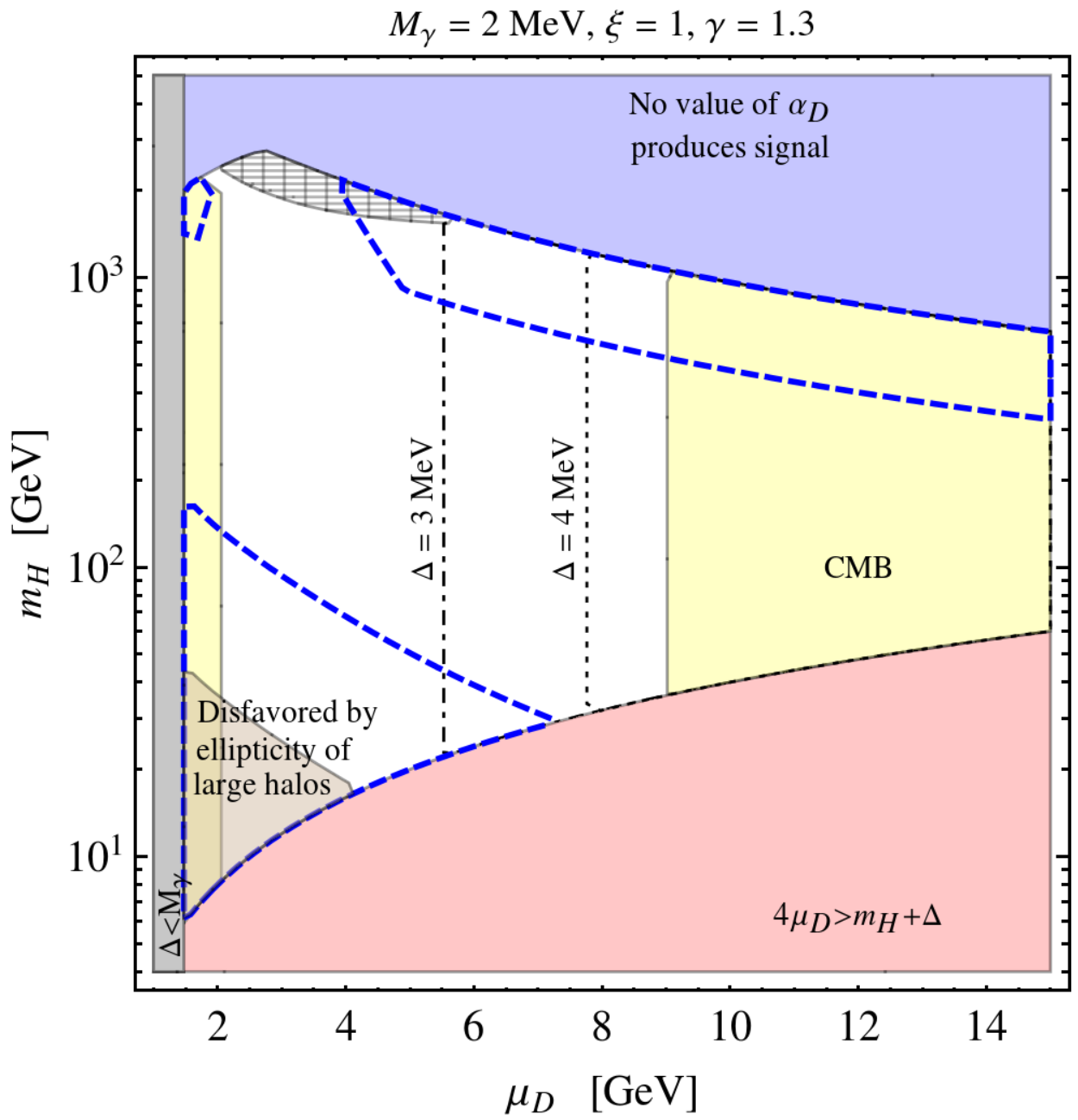}
\caption{Same as in Fig.~\ref{fig:param_space_with_constraints_partial} (partially ionized dark matter), for a contracted NFW profile with $\gamma = 1.3$. }
\label{fig:param_space_with_constraints_2_partial}
\end{figure}

Having outlined the region of parameter space in which the observed 511~keV line can be produced, we now consider the relevant astrophysical and cosmological constraints, and present them in Figs.~\ref{fig:param_space_with_constraints} and \ref{fig:param_space_with_constraints_2}, for the fully ionized branch, and Figs.~\ref{fig:param_space_with_constraints_partial} and \ref{fig:param_space_with_constraints_2_partial} for the partially ionized branch.

The most stringent constraint comes from CMB data.  Bound state formation during recombination deposits additional energy into the CMB, which may cause distortions~\cite{Madhavacheril:2013cna}.  The rate of energy deposition from bound state formation is
\begin{equation}
\dfrac{dE}{dV \, dt} = \rho_c^2 \, \Omega\DM^2 (1 + z)^6 p\BSF(z),
\end{equation}
where $\rho_c$ is the critical density of the universe, $\Omega\DM\simeq0.26$, $z$ is the redshift, and
\begin{equation}
p\BSF(z) = f(z) \, s\BSF \, \Delta \, .
\label{eq:pBSF}
\end{equation}
Here, $f(z)$ is a (redshift-dependent) factor which describes the fraction of energy absorbed by the CMB; for the two-stage process of dark photon emission and decay into $e^+e^-$, $f(z) \simeq 0.45$~\cite{Madhavacheril:2013cna}. Note that in Eq.~\eqref{eq:pBSF} we have used $s\BSF$, which incorporates the (possibly non-maximal) residual ionization fraction of DM, since dark recombination is expected to precede ordinary recombination. We evaluate $s\BSF$ by replacing $\vrel \to \Mg/\muD$, noting that the non-zero dark photon mass imposes a cutoff on the Sommerfeld enhancement of $(\sigma\vrel)\BSF$, as discussed below Eq.~\eqref{eq:sigmav_approx}. At the time of CMB, the $\pD-\eD$ relative velocity would be $\vrel \approx \sqrt{3\xi \,T_{_{\rm CMB}}/\muD}$, were the dark ions in equilibrium with the dark photons; however, $\vrel$ is in fact significantly lower, since the dark ions have already decoupled from the dark photons~\cite{CyrRacine:2012fz}. For $\Mg \gtrsim 2~\MeV$, the Sommerfeld enhancement of $(\sigma \vrel)\BSF$ has indeed saturated. We require~\cite{Madhavacheril:2013cna} 
\begin{equation}
p\BSF < 0.66 \times 10^{-6} \ \frac{\rm m^3}{\rm s \; kg} \, .
\end{equation}

The spectrum of the injected positrons from the decay of the dark photons emitted in the formation of dark atoms, is flat and extends between the energies
\begin{align}
\dfrac{\omega_\gamma}{2} 
\left[1 \mp \sqrt{ (1 - \Mg^2 / \omega_\gamma^2 ) (1 - 4 \tilde{m}_e^2 / \Mg^2 ) } \right] \, ,
\end{align}
where $\omega_\gamma \simeq \Delta$ is the energy of the dark photon. We require $\Delta < 20~\MeV$, such that the average injection energy of the positrons is $\lesssim 10~\MeV$. This is consistent with the constraints derived from the diffuse $\gamma$-ray flux discussed in Sec.~\ref{sec:511 keV - general}, particularly considering the non-monoenergetic positron injection spectrum. However, in the parameter space which can account for the 511~keV flux and satisfies the CMB constraints, we find that $\Delta$ is even smaller.

The DM self-scattering inside haloes may affect the dynamics of galaxies. In smaller haloes, the effect of the DM self-interactions can bring theory in better agreement with observations. This parameter space is denoted by the blue dashed line, which encloses the region in which the cross section is greater than $0.5 \; \mathrm{cm}^2 \slash \mathrm{g}$ at $10 \; \mathrm{km} \slash \mathrm{s}$.  Currently, the most stringent constraint on the DM self-interactions is thought to arise from the observed ellipticity of galaxies of the size of the Milky Way~\cite{Rocha:2012jg,Peter:2012jh}.  We have shaded brown the region where $\sigma_\mathrm{scat} \slash m_\mathrm{DM} > 1 \; \mathrm{cm}^2 \slash \mathrm{g}$ at $220 \; \mathrm{km} \slash \mathrm{s}$.  This follows the analysis of~\cite{Petraki:2014uza}.

Moreover, we require that the dark force provides efficient annihilation of DM in the early universe, with the lower bound on $\aD$ given in Eq.~\eqref{eq:alpha ann}. This does depend weakly on $\xi_{\mathrm{FO}}$.  We note that the decoupling of the dark and visible sectors depends on the kinetic mixing $\epsilon$ and dark electron mass; however, if the two sectors decoupled kinetically before DM freeze-out, then $\xi_\mathrm{FO}$ will not generally equal $\xi_\mathrm{DR}$.  However, $\xi$ is slowly varying and the $\xi$ dependence of this constraint is rather mild.  Therefore, for simplicity, we have taken $\xi_\mathrm{FO} = \xi_\mathrm{DR}$.

In the entire parameter space of interest, $\aD$ is well below the unitarity bound of Eq.~\eqref{eq:alpha uni}. We enforce the consistency condition of Eq.~\eqref{eq:consistency}, and we exclude the regions where $\Mg > \Delta$, since dark atoms could not then form via emission of a dark photon.

Successful BBN constrains the radiation present in the universe at temperatures $T_V = T_{\rm BBN} \sim 1~\MeV$. A relativistic dark plasma consisting of dark photons, present at BBN, would need to be at temperature $T_D \lesssim 0.6 \, T_V$. Dark photons with mass $\Mg \gtrsim 2~\MeV$ would be quasi-relativistic or non-relativistic at BBN and this constraint ($\xi \lesssim 0.6$) is at least somewhat relaxed. Nevertheless, to retain $\xi<1$, it is necessary to ensure that the dark and the ordinary sectors do not equilibrate via the kinetic mixing $\epsilon$. This poses the condition $\epsilon^2\aD \lesssim 10^{-19}$~\cite{Petraki:2014uza}, where we assumed that the lightest charged particle in the dark sector is the dark Higgs which gives mass to the dark photon, and has itself only a somewhat larger mass than the dark photon. (If the dark photon acquires its mass via the St\"{u}ckelberg mechanism, then $\eD$ is the lightest dark charged particle, and the above constraint is significantly relaxed.) Direct detection experiments suggest a similar upper bound on $\epsilon$. (See e.g.~Ref.~\cite{Fornengo:2011sz,Kaplinghat:2013yxa}, although the bounds are somewhat strengthened by more recent experimental data~\cite{Akerib:2013tjd}. However, a dedicated analysis of direct detection data for atomic DM, which would take into account the different DM elements and the various possible DM-nucleon interactions, is needed.) 
These upper bounds on $\epsilon$ are very comfortably compatible with the lower bound estimated in Eq.~\eqref{eq:epsilon min}, which ensures prompt decay of the dark photons inside the halo.

\bigskip
\section{Conclusion \label{sec:Conc}}

The asymmetric DM scenario allows for rich dark sector microphysics, which can yield distinct detection signatures. 
Because in this scenario, the DM relic abundance is protected by a conserved particle-number excess, the observed DM density does not set an upper bound on the DM couplings to lighter species, despite DM being a thermal relic. Dark matter may then possess significant couplings to light force mediators; this is, in fact, motivated by the observed galactic structure, which currently can be explained better by DM with sizable self-interactions rather than by collisionless cold DM. Similarly to ordinary matter, DM with long-range interactions may possess rich phenomenology. The formation of bound states in particular, is an important feature of many such theories.

In this paper, we have explored indirect detection signals arising from the formation of atomic bound states in haloes today, by asymmetric DM which couples to a light dark photon. Level transitions, such as bound-state formation, can yield low-energy signals which cannot otherwise be easily produced by annihilating thermal-relic DM. However, level transitions may also produce signals in the spectrum expected from WIMP annihilation. In our analysis, we explored the expected signal strength from dark atom formation generically. We also showed that the radiative formation of dark atoms can account for the observed 511 keV line, provided that the DM profile in the central kpc of our galaxy is steep. The observed radiation backgrounds can constrain the parameter space of atomic DM with kinetic mixing to hypercharge, based on the radiation emitted from the formation of bound states, and so also allow correlation of direct and indirect detection bounds.

\section*{Acknowledgements}

A.K. was supported by DOE Grant DE-SC0009937 and by the World Premier International Research Center Initiative (WPI Initiative), MEXT, Japan.  K.P. was supported by the Netherlands Foundation for Fundamental Research of Matter (FOM) and the Netherlands Organisation for Scientific Research (NWO).

\bibliography{Bibliography.bib}

\begin{thebibliography}{78}%
\makeatletter
\providecommand \@ifxundefined [1]{%
 \@ifx{#1\undefined}
}%
\providecommand \@ifnum [1]{%
 \ifnum #1\expandafter \@firstoftwo
 \else \expandafter \@secondoftwo
 \fi
}%
\providecommand \@ifx [1]{%
 \ifx #1\expandafter \@firstoftwo
 \else \expandafter \@secondoftwo
 \fi
}%
\providecommand \natexlab [1]{#1}%
\providecommand \enquote  [1]{``#1''}%
\providecommand \bibnamefont  [1]{#1}%
\providecommand \bibfnamefont [1]{#1}%
\providecommand \citenamefont [1]{#1}%
\providecommand \href@noop [0]{\@secondoftwo}%
\providecommand \href [0]{\begingroup \@sanitize@url \@href}%
\providecommand \@href[1]{\@@startlink{#1}\@@href}%
\providecommand \@@href[1]{\endgroup#1\@@endlink}%
\providecommand \@sanitize@url [0]{\catcode `\\12\catcode `\$12\catcode
  `\&12\catcode `\#12\catcode `\^12\catcode `\_12\catcode `\%12\relax}%
\providecommand \@@startlink[1]{}%
\providecommand \@@endlink[0]{}%
\providecommand \url  [0]{\begingroup\@sanitize@url \@url }%
\providecommand \@url [1]{\endgroup\@href {#1}{\urlprefix }}%
\providecommand \urlprefix  [0]{URL }%
\providecommand \Eprint [0]{\href }%
\providecommand \doibase [0]{http://dx.doi.org/}%
\providecommand \selectlanguage [0]{\@gobble}%
\providecommand \bibinfo  [0]{\@secondoftwo}%
\providecommand \bibfield  [0]{\@secondoftwo}%
\providecommand \translation [1]{[#1]}%
\providecommand \BibitemOpen [0]{}%
\providecommand \bibitemStop [0]{}%
\providecommand \bibitemNoStop [0]{.\EOS\space}%
\providecommand \EOS [0]{\spacefactor3000\relax}%
\providecommand \BibitemShut  [1]{\csname bibitem#1\endcsname}%
\let\auto@bib@innerbib\@empty
\bibitem [{\citenamefont {Kaplan}\ \emph {et~al.}(2010)\citenamefont {Kaplan},
  \citenamefont {Krnjaic}, \citenamefont {Rehermann},\ and\ \citenamefont
  {Wells}}]{Kaplan:2009de}%
  \BibitemOpen
  \bibfield  {author} {\bibinfo {author} {\bibfnamefont {D.~E.}\ \bibnamefont
  {Kaplan}}, \bibinfo {author} {\bibfnamefont {G.~Z.}\ \bibnamefont {Krnjaic}},
  \bibinfo {author} {\bibfnamefont {K.~R.}\ \bibnamefont {Rehermann}}, \ and\
  \bibinfo {author} {\bibfnamefont {C.~M.}\ \bibnamefont {Wells}},\ }\href
  {\doibase 10.1088/1475-7516} {\bibfield  {journal} {\bibinfo  {journal}
  {JCAP}\ }\textbf {\bibinfo {volume} {1005}},\ \bibinfo {pages} {021}
  (\bibinfo {year} {2010})},\ \Eprint {http://arxiv.org/abs/0909.0753}
  {arXiv:0909.0753 [hep-ph]} \BibitemShut {NoStop}%
\bibitem [{\citenamefont {Kaplan}\ \emph {et~al.}(2011)\citenamefont {Kaplan},
  \citenamefont {Krnjaic}, \citenamefont {Rehermann},\ and\ \citenamefont
  {Wells}}]{Kaplan:2011yj}%
  \BibitemOpen
  \bibfield  {author} {\bibinfo {author} {\bibfnamefont {D.~E.}\ \bibnamefont
  {Kaplan}}, \bibinfo {author} {\bibfnamefont {G.~Z.}\ \bibnamefont {Krnjaic}},
  \bibinfo {author} {\bibfnamefont {K.~R.}\ \bibnamefont {Rehermann}}, \ and\
  \bibinfo {author} {\bibfnamefont {C.~M.}\ \bibnamefont {Wells}},\ }\href
  {\doibase 10.1088/1475-7516} {\bibfield  {journal} {\bibinfo  {journal}
  {JCAP}\ }\textbf {\bibinfo {volume} {1110}},\ \bibinfo {pages} {011}
  (\bibinfo {year} {2011})},\ \Eprint {http://arxiv.org/abs/1105.2073}
  {arXiv:1105.2073 [hep-ph]} \BibitemShut {NoStop}%
\bibitem [{\citenamefont {Petraki}\ \emph {et~al.}(2012)\citenamefont
  {Petraki}, \citenamefont {Trodden},\ and\ \citenamefont
  {Volkas}}]{Petraki:2011mv}%
  \BibitemOpen
  \bibfield  {author} {\bibinfo {author} {\bibfnamefont {K.}~\bibnamefont
  {Petraki}}, \bibinfo {author} {\bibfnamefont {M.}~\bibnamefont {Trodden}}, \
  and\ \bibinfo {author} {\bibfnamefont {R.~R.}\ \bibnamefont {Volkas}},\
  }\href {\doibase 10.1088/1475-7516} {\bibfield  {journal} {\bibinfo
  {journal} {JCAP}\ }\textbf {\bibinfo {volume} {1202}},\ \bibinfo {pages}
  {044} (\bibinfo {year} {2012})},\ \Eprint {http://arxiv.org/abs/1111.4786}
  {arXiv:1111.4786 [hep-ph]} \BibitemShut {NoStop}%
\bibitem [{\citenamefont {{von Harling}}\ \emph {et~al.}(2012)\citenamefont
  {{von Harling}}, \citenamefont {Petraki},\ and\ \citenamefont
  {Volkas}}]{vonHarling:2012yn}%
  \BibitemOpen
  \bibfield  {author} {\bibinfo {author} {\bibfnamefont {B.}~\bibnamefont {{von
  Harling}}}, \bibinfo {author} {\bibfnamefont {K.}~\bibnamefont {Petraki}}, \
  and\ \bibinfo {author} {\bibfnamefont {R.~R.}\ \bibnamefont {Volkas}},\
  }\href {\doibase 10.1088/1475-7516} {\bibfield  {journal} {\bibinfo
  {journal} {JCAP}\ }\textbf {\bibinfo {volume} {1205}},\ \bibinfo {pages}
  {021} (\bibinfo {year} {2012})},\ \Eprint {http://arxiv.org/abs/1201.2200}
  {arXiv:1201.2200 [hep-ph]} \BibitemShut {NoStop}%
\bibitem [{\citenamefont {Davoudiasl}\ and\ \citenamefont
  {Mohapatra}(2012)}]{Davoudiasl:2012uw}%
  \BibitemOpen
  \bibfield  {author} {\bibinfo {author} {\bibfnamefont {H.}~\bibnamefont
  {Davoudiasl}}\ and\ \bibinfo {author} {\bibfnamefont {R.~N.}\ \bibnamefont
  {Mohapatra}},\ }\href {\doibase 10.1088/1367-2630} {\bibfield  {journal}
  {\bibinfo  {journal} {New J.Phys.}\ }\textbf {\bibinfo {volume} {14}},\
  \bibinfo {pages} {095011} (\bibinfo {year} {2012})},\ \Eprint
  {http://arxiv.org/abs/1203.1247} {arXiv:1203.1247 [hep-ph]} \BibitemShut
  {NoStop}%
\bibitem [{\citenamefont {Petraki}\ and\ \citenamefont
  {Volkas}(2013)}]{Petraki:2013wwa}%
  \BibitemOpen
  \bibfield  {author} {\bibinfo {author} {\bibfnamefont {K.}~\bibnamefont
  {Petraki}}\ and\ \bibinfo {author} {\bibfnamefont {R.~R.}\ \bibnamefont
  {Volkas}},\ }\href {\doibase 10.1142/S0217751X13300287} {\bibfield  {journal}
  {\bibinfo  {journal} {Int.J.Mod.Phys.}\ }\textbf {\bibinfo {volume} {A28}},\
  \bibinfo {pages} {1330028} (\bibinfo {year} {2013})},\ \Eprint
  {http://arxiv.org/abs/1305.4939} {arXiv:1305.4939 [hep-ph]} \BibitemShut
  {NoStop}%
\bibitem [{\citenamefont {Zurek}(2014)}]{Zurek:2013wia}%
  \BibitemOpen
  \bibfield  {author} {\bibinfo {author} {\bibfnamefont {K.~M.}\ \bibnamefont
  {Zurek}},\ }\href {\doibase 10.1016/j.physrep.2013.12.001} {\bibfield
  {journal} {\bibinfo  {journal} {Phys.Rept.}\ }\textbf {\bibinfo {volume}
  {537}},\ \bibinfo {pages} {91} (\bibinfo {year} {2014})},\ \Eprint
  {http://arxiv.org/abs/1308.0338} {arXiv:1308.0338 [hep-ph]} \BibitemShut
  {NoStop}%
\bibitem [{\citenamefont {Boucenna}\ and\ \citenamefont
  {Morisi}(2014)}]{Boucenna:2013wba}%
  \BibitemOpen
  \bibfield  {author} {\bibinfo {author} {\bibfnamefont {S.}~\bibnamefont
  {Boucenna}}\ and\ \bibinfo {author} {\bibfnamefont {S.}~\bibnamefont
  {Morisi}},\ }\href {\doibase 10.3389/fphy.2013.00033} {\bibfield  {journal}
  {\bibinfo  {journal} {Front.Phys.}\ }\textbf {\bibinfo {volume} {1}},\
  \bibinfo {pages} {33} (\bibinfo {year} {2014})},\ \Eprint
  {http://arxiv.org/abs/1310.1904} {arXiv:1310.1904 [hep-ph]} \BibitemShut
  {NoStop}%
\bibitem [{\citenamefont {Spergel}\ and\ \citenamefont
  {Steinhardt}(2000)}]{Spergel:1999mh}%
  \BibitemOpen
  \bibfield  {author} {\bibinfo {author} {\bibfnamefont {D.~N.}\ \bibnamefont
  {Spergel}}\ and\ \bibinfo {author} {\bibfnamefont {P.~J.}\ \bibnamefont
  {Steinhardt}},\ }\href {\doibase 10.1103/PhysRevLett.84.3760} {\bibfield
  {journal} {\bibinfo  {journal} {Phys.Rev.Lett.}\ }\textbf {\bibinfo {volume}
  {84}},\ \bibinfo {pages} {3760} (\bibinfo {year} {2000})},\ \Eprint
  {http://arxiv.org/abs/astro-ph/9909386} {arXiv:astro-ph/9909386 [astro-ph]}
  \BibitemShut {NoStop}%
\bibitem [{\citenamefont {Wandelt}\ \emph {et~al.}(2000)\citenamefont
  {Wandelt}, \citenamefont {Dave}, \citenamefont {Farrar}, \citenamefont
  {McGuire}, \citenamefont {Spergel} \emph {et~al.}}]{Wandelt:2000ad}%
  \BibitemOpen
  \bibfield  {author} {\bibinfo {author} {\bibfnamefont {B.~D.}\ \bibnamefont
  {Wandelt}}, \bibinfo {author} {\bibfnamefont {R.}~\bibnamefont {Dave}},
  \bibinfo {author} {\bibfnamefont {G.~R.}\ \bibnamefont {Farrar}}, \bibinfo
  {author} {\bibfnamefont {P.~C.}\ \bibnamefont {McGuire}}, \bibinfo {author}
  {\bibfnamefont {D.~N.}\ \bibnamefont {Spergel}},  \emph {et~al.},\
  }\href@noop {} {\ ,\ \bibinfo {pages} {263} (\bibinfo {year} {2000})},\
  \Eprint {http://arxiv.org/abs/astro-ph/0006344} {arXiv:astro-ph/0006344
  [astro-ph]} \BibitemShut {NoStop}%
\bibitem [{\citenamefont {Faraggi}\ and\ \citenamefont
  {Pospelov}(2002)}]{Faraggi:2000pv}%
  \BibitemOpen
  \bibfield  {author} {\bibinfo {author} {\bibfnamefont {A.~E.}\ \bibnamefont
  {Faraggi}}\ and\ \bibinfo {author} {\bibfnamefont {M.}~\bibnamefont
  {Pospelov}},\ }\href {\doibase 10.1016/S0927-6505(01)00121-9} {\bibfield
  {journal} {\bibinfo  {journal} {Astropart.Phys.}\ }\textbf {\bibinfo {volume}
  {16}},\ \bibinfo {pages} {451} (\bibinfo {year} {2002})},\ \Eprint
  {http://arxiv.org/abs/hep-ph/0008223} {arXiv:hep-ph/0008223 [hep-ph]}
  \BibitemShut {NoStop}%
\bibitem [{\citenamefont {Kusenko}\ and\ \citenamefont
  {Steinhardt}(2001)}]{Kusenko:2001vu}%
  \BibitemOpen
  \bibfield  {author} {\bibinfo {author} {\bibfnamefont {A.}~\bibnamefont
  {Kusenko}}\ and\ \bibinfo {author} {\bibfnamefont {P.~J.}\ \bibnamefont
  {Steinhardt}},\ }\href {\doibase 10.1103/PhysRevLett.87.141301} {\bibfield
  {journal} {\bibinfo  {journal} {Phys.Rev.Lett.}\ }\textbf {\bibinfo {volume}
  {87}},\ \bibinfo {pages} {141301} (\bibinfo {year} {2001})},\ \Eprint
  {http://arxiv.org/abs/astro-ph/0106008} {arXiv:astro-ph/0106008 [astro-ph]}
  \BibitemShut {NoStop}%
\bibitem [{\citenamefont {Mohapatra}\ \emph {et~al.}(2002)\citenamefont
  {Mohapatra}, \citenamefont {Nussinov},\ and\ \citenamefont
  {Teplitz}}]{Mohapatra:2001sx}%
  \BibitemOpen
  \bibfield  {author} {\bibinfo {author} {\bibfnamefont {R.}~\bibnamefont
  {Mohapatra}}, \bibinfo {author} {\bibfnamefont {S.}~\bibnamefont {Nussinov}},
  \ and\ \bibinfo {author} {\bibfnamefont {V.}~\bibnamefont {Teplitz}},\ }\href
  {\doibase 10.1103/PhysRevD.66.063002} {\bibfield  {journal} {\bibinfo
  {journal} {Phys.Rev.}\ }\textbf {\bibinfo {volume} {D66}},\ \bibinfo {pages}
  {063002} (\bibinfo {year} {2002})},\ \Eprint
  {http://arxiv.org/abs/hep-ph/0111381} {arXiv:hep-ph/0111381 [hep-ph]}
  \BibitemShut {NoStop}%
\bibitem [{\citenamefont {Feng}\ \emph {et~al.}(2009)\citenamefont {Feng},
  \citenamefont {Kaplinghat}, \citenamefont {Tu},\ and\ \citenamefont
  {Yu}}]{Feng:2009mn}%
  \BibitemOpen
  \bibfield  {author} {\bibinfo {author} {\bibfnamefont {J.~L.}\ \bibnamefont
  {Feng}}, \bibinfo {author} {\bibfnamefont {M.}~\bibnamefont {Kaplinghat}},
  \bibinfo {author} {\bibfnamefont {H.}~\bibnamefont {Tu}}, \ and\ \bibinfo
  {author} {\bibfnamefont {H.-B.}\ \bibnamefont {Yu}},\ }\href {\doibase
  10.1088/1475-7516} {\bibfield  {journal} {\bibinfo  {journal} {JCAP}\
  }\textbf {\bibinfo {volume} {0907}},\ \bibinfo {pages} {004} (\bibinfo {year}
  {2009})},\ \Eprint {http://arxiv.org/abs/0905.3039} {arXiv:0905.3039
  [hep-ph]} \BibitemShut {NoStop}%
\bibitem [{\citenamefont {Feng}\ \emph {et~al.}(2010)\citenamefont {Feng},
  \citenamefont {Kaplinghat},\ and\ \citenamefont {Yu}}]{Feng:2009hw}%
  \BibitemOpen
  \bibfield  {author} {\bibinfo {author} {\bibfnamefont {J.~L.}\ \bibnamefont
  {Feng}}, \bibinfo {author} {\bibfnamefont {M.}~\bibnamefont {Kaplinghat}}, \
  and\ \bibinfo {author} {\bibfnamefont {H.-B.}\ \bibnamefont {Yu}},\ }\href
  {\doibase 10.1103/PhysRevLett.104.151301} {\bibfield  {journal} {\bibinfo
  {journal} {Phys.Rev.Lett.}\ }\textbf {\bibinfo {volume} {104}},\ \bibinfo
  {pages} {151301} (\bibinfo {year} {2010})},\ \Eprint
  {http://arxiv.org/abs/0911.0422} {arXiv:0911.0422 [hep-ph]} \BibitemShut
  {NoStop}%
\bibitem [{\citenamefont {Loeb}\ and\ \citenamefont
  {Weiner}(2011)}]{Loeb:2010gj}%
  \BibitemOpen
  \bibfield  {author} {\bibinfo {author} {\bibfnamefont {A.}~\bibnamefont
  {Loeb}}\ and\ \bibinfo {author} {\bibfnamefont {N.}~\bibnamefont {Weiner}},\
  }\href {\doibase 10.1103/PhysRevLett.106.171302} {\bibfield  {journal}
  {\bibinfo  {journal} {Phys.Rev.Lett.}\ }\textbf {\bibinfo {volume} {106}},\
  \bibinfo {pages} {171302} (\bibinfo {year} {2011})},\ \Eprint
  {http://arxiv.org/abs/1011.6374} {arXiv:1011.6374 [astro-ph.CO]} \BibitemShut
  {NoStop}%
\bibitem [{\citenamefont {{Spier Moreira Alves}}\ \emph
  {et~al.}(2010)\citenamefont {{Spier Moreira Alves}}, \citenamefont
  {Behbahani}, \citenamefont {Schuster},\ and\ \citenamefont
  {Wacker}}]{Alves:2010dd}%
  \BibitemOpen
  \bibfield  {author} {\bibinfo {author} {\bibfnamefont {D.}~\bibnamefont
  {{Spier Moreira Alves}}}, \bibinfo {author} {\bibfnamefont {S.~R.}\
  \bibnamefont {Behbahani}}, \bibinfo {author} {\bibfnamefont {P.}~\bibnamefont
  {Schuster}}, \ and\ \bibinfo {author} {\bibfnamefont {J.~G.}\ \bibnamefont
  {Wacker}},\ }\href {\doibase 10.1007/JHEP06(2010)113} {\bibfield  {journal}
  {\bibinfo  {journal} {JHEP}\ }\textbf {\bibinfo {volume} {1006}},\ \bibinfo
  {pages} {113} (\bibinfo {year} {2010})},\ \Eprint
  {http://arxiv.org/abs/1003.4729} {arXiv:1003.4729 [hep-ph]} \BibitemShut
  {NoStop}%
\bibitem [{\citenamefont {Rocha}\ \emph {et~al.}(2013)\citenamefont {Rocha},
  \citenamefont {Peter}, \citenamefont {Bullock}, \citenamefont {Kaplinghat},
  \citenamefont {Garrison-Kimmel} \emph {et~al.}}]{Rocha:2012jg}%
  \BibitemOpen
  \bibfield  {author} {\bibinfo {author} {\bibfnamefont {M.}~\bibnamefont
  {Rocha}}, \bibinfo {author} {\bibfnamefont {A.~H.}\ \bibnamefont {Peter}},
  \bibinfo {author} {\bibfnamefont {J.~S.}\ \bibnamefont {Bullock}}, \bibinfo
  {author} {\bibfnamefont {M.}~\bibnamefont {Kaplinghat}}, \bibinfo {author}
  {\bibfnamefont {S.}~\bibnamefont {Garrison-Kimmel}},  \emph {et~al.},\ }\href
  {\doibase 10.1093/mnras} {\bibfield  {journal} {\bibinfo  {journal}
  {Mon.Not.Roy.Astron.Soc.}\ }\textbf {\bibinfo {volume} {430}},\ \bibinfo
  {pages} {81} (\bibinfo {year} {2013})},\ \Eprint
  {http://arxiv.org/abs/1208.3025} {arXiv:1208.3025 [astro-ph.CO]} \BibitemShut
  {NoStop}%
\bibitem [{\citenamefont {Peter}\ \emph {et~al.}(2013)\citenamefont {Peter},
  \citenamefont {Rocha}, \citenamefont {Bullock},\ and\ \citenamefont
  {Kaplinghat}}]{Peter:2012jh}%
  \BibitemOpen
  \bibfield  {author} {\bibinfo {author} {\bibfnamefont {A.~H.}\ \bibnamefont
  {Peter}}, \bibinfo {author} {\bibfnamefont {M.}~\bibnamefont {Rocha}},
  \bibinfo {author} {\bibfnamefont {J.~S.}\ \bibnamefont {Bullock}}, \ and\
  \bibinfo {author} {\bibfnamefont {M.}~\bibnamefont {Kaplinghat}},\ }\href
  {\doibase 10.1093/mnras/sts535} {\bibfield  {journal} {\bibinfo  {journal}
  {Mon.Not.Roy.Astron.Soc.}\ }\textbf {\bibinfo {volume} {430}},\ \bibinfo
  {pages} {105} (\bibinfo {year} {2013})},\ \Eprint
  {http://arxiv.org/abs/1208.3026} {arXiv:1208.3026 [astro-ph.CO]} \BibitemShut
  {NoStop}%
\bibitem [{\citenamefont {Vogelsberger}\ \emph {et~al.}(2012)\citenamefont
  {Vogelsberger}, \citenamefont {Zavala},\ and\ \citenamefont
  {Loeb}}]{Vogelsberger:2012ku}%
  \BibitemOpen
  \bibfield  {author} {\bibinfo {author} {\bibfnamefont {M.}~\bibnamefont
  {Vogelsberger}}, \bibinfo {author} {\bibfnamefont {J.}~\bibnamefont
  {Zavala}}, \ and\ \bibinfo {author} {\bibfnamefont {A.}~\bibnamefont
  {Loeb}},\ }\href@noop {} {\bibfield  {journal} {\bibinfo  {journal}
  {Mon.Not.Roy.Astron.Soc.}\ }\textbf {\bibinfo {volume} {423}},\ \bibinfo
  {pages} {3740} (\bibinfo {year} {2012})},\ \Eprint
  {http://arxiv.org/abs/1201.5892} {arXiv:1201.5892 [astro-ph.CO]} \BibitemShut
  {NoStop}%
\bibitem [{\citenamefont {Vogelsberger}\ and\ \citenamefont
  {Zavala}(2013)}]{Vogelsberger:2012sa}%
  \BibitemOpen
  \bibfield  {author} {\bibinfo {author} {\bibfnamefont {M.}~\bibnamefont
  {Vogelsberger}}\ and\ \bibinfo {author} {\bibfnamefont {J.}~\bibnamefont
  {Zavala}},\ }\href {\doibase 10.1093/mnras} {\bibfield  {journal} {\bibinfo
  {journal} {Mon.Not.Roy.Astron.Soc.}\ }\textbf {\bibinfo {volume} {430}},\
  \bibinfo {pages} {1722} (\bibinfo {year} {2013})},\ \Eprint
  {http://arxiv.org/abs/1211.1377} {arXiv:1211.1377 [astro-ph.CO]} \BibitemShut
  {NoStop}%
\bibitem [{\citenamefont {Zavala}\ \emph {et~al.}(2013)\citenamefont {Zavala},
  \citenamefont {Vogelsberger},\ and\ \citenamefont {Walker}}]{Zavala:2012us}%
  \BibitemOpen
  \bibfield  {author} {\bibinfo {author} {\bibfnamefont {J.}~\bibnamefont
  {Zavala}}, \bibinfo {author} {\bibfnamefont {M.}~\bibnamefont
  {Vogelsberger}}, \ and\ \bibinfo {author} {\bibfnamefont {M.~G.}\
  \bibnamefont {Walker}},\ }\href {\doibase 10.1093/mnrasl} {\bibfield
  {journal} {\bibinfo  {journal} {Monthly Notices of the Royal Astronomical
  Society: Letters}\ }\textbf {\bibinfo {volume} {431}},\ \bibinfo {pages}
  {L20} (\bibinfo {year} {2013})},\ \Eprint {http://arxiv.org/abs/1211.6426}
  {arXiv:1211.6426 [astro-ph.CO]} \BibitemShut {NoStop}%
\bibitem [{\citenamefont {Cline}\ \emph {et~al.}(2012)\citenamefont {Cline},
  \citenamefont {Liu},\ and\ \citenamefont {Xue}}]{Cline:2012is}%
  \BibitemOpen
  \bibfield  {author} {\bibinfo {author} {\bibfnamefont {J.~M.}\ \bibnamefont
  {Cline}}, \bibinfo {author} {\bibfnamefont {Z.}~\bibnamefont {Liu}}, \ and\
  \bibinfo {author} {\bibfnamefont {W.}~\bibnamefont {Xue}},\ }\href {\doibase
  10.1103/PhysRevD.85.101302} {\bibfield  {journal} {\bibinfo  {journal}
  {Phys.Rev.}\ }\textbf {\bibinfo {volume} {D85}},\ \bibinfo {pages} {101302}
  (\bibinfo {year} {2012})},\ \Eprint {http://arxiv.org/abs/1201.4858}
  {arXiv:1201.4858 [hep-ph]} \BibitemShut {NoStop}%
\bibitem [{\citenamefont {Cyr-Racine}\ \emph {et~al.}(2014)\citenamefont
  {Cyr-Racine}, \citenamefont {{de Putter}}, \citenamefont {Raccanelli},\ and\
  \citenamefont {Sigurdson}}]{Cyr-Racine:2013fsa}%
  \BibitemOpen
  \bibfield  {author} {\bibinfo {author} {\bibfnamefont {F.-Y.}\ \bibnamefont
  {Cyr-Racine}}, \bibinfo {author} {\bibfnamefont {R.}~\bibnamefont {{de
  Putter}}}, \bibinfo {author} {\bibfnamefont {A.}~\bibnamefont {Raccanelli}},
  \ and\ \bibinfo {author} {\bibfnamefont {K.}~\bibnamefont {Sigurdson}},\
  }\href {\doibase 10.1103/PhysRevD.89.063517} {\bibfield  {journal} {\bibinfo
  {journal} {Phys.Rev.}\ }\textbf {\bibinfo {volume} {D89}},\ \bibinfo {pages}
  {063517} (\bibinfo {year} {2014})},\ \Eprint {http://arxiv.org/abs/1310.3278}
  {arXiv:1310.3278 [astro-ph.CO]} \BibitemShut {NoStop}%
\bibitem [{\citenamefont {Cyr-Racine}\ and\ \citenamefont
  {Sigurdson}(2013)}]{CyrRacine:2012fz}%
  \BibitemOpen
  \bibfield  {author} {\bibinfo {author} {\bibfnamefont {F.-Y.}\ \bibnamefont
  {Cyr-Racine}}\ and\ \bibinfo {author} {\bibfnamefont {K.}~\bibnamefont
  {Sigurdson}},\ }\href {\doibase 10.1103/PhysRevD.87.103515} {\bibfield
  {journal} {\bibinfo  {journal} {Phys.Rev.}\ }\textbf {\bibinfo {volume}
  {D87}},\ \bibinfo {pages} {103515} (\bibinfo {year} {2013})},\ \Eprint
  {http://arxiv.org/abs/1209.5752} {arXiv:1209.5752 [astro-ph.CO]} \BibitemShut
  {NoStop}%
\bibitem [{\citenamefont {Cline}\ \emph
  {et~al.}(2014{\natexlab{a}})\citenamefont {Cline}, \citenamefont {Liu},
  \citenamefont {Moore},\ and\ \citenamefont {Xue}}]{Cline:2013pca}%
  \BibitemOpen
  \bibfield  {author} {\bibinfo {author} {\bibfnamefont {J.~M.}\ \bibnamefont
  {Cline}}, \bibinfo {author} {\bibfnamefont {Z.}~\bibnamefont {Liu}}, \bibinfo
  {author} {\bibfnamefont {G.}~\bibnamefont {Moore}}, \ and\ \bibinfo {author}
  {\bibfnamefont {W.}~\bibnamefont {Xue}},\ }\href {\doibase
  10.1103/PhysRevD.89.043514} {\bibfield  {journal} {\bibinfo  {journal}
  {Phys.Rev.}\ }\textbf {\bibinfo {volume} {D89}},\ \bibinfo {pages} {043514}
  (\bibinfo {year} {2014}{\natexlab{a}})},\ \Eprint
  {http://arxiv.org/abs/1311.6468} {arXiv:1311.6468 [hep-ph]} \BibitemShut
  {NoStop}%
\bibitem [{\citenamefont {Petraki}\ \emph {et~al.}(2014)\citenamefont
  {Petraki}, \citenamefont {Pearce},\ and\ \citenamefont
  {Kusenko}}]{Petraki:2014uza}%
  \BibitemOpen
  \bibfield  {author} {\bibinfo {author} {\bibfnamefont {K.}~\bibnamefont
  {Petraki}}, \bibinfo {author} {\bibfnamefont {L.}~\bibnamefont {Pearce}}, \
  and\ \bibinfo {author} {\bibfnamefont {A.}~\bibnamefont {Kusenko}},\ }\href
  {\doibase 10.1088/1475-7516/2014/07/039} {\bibfield  {journal} {\bibinfo
  {journal} {JCAP}\ }\textbf {\bibinfo {volume} {1407}},\ \bibinfo {pages}
  {039} (\bibinfo {year} {2014})},\ \Eprint {http://arxiv.org/abs/1403.1077}
  {arXiv:1403.1077 [hep-ph]} \BibitemShut {NoStop}%
\bibitem [{\citenamefont {Foot}\ and\ \citenamefont
  {Vagnozzi}(2015)}]{Foot:2014uba}%
  \BibitemOpen
  \bibfield  {author} {\bibinfo {author} {\bibfnamefont {R.}~\bibnamefont
  {Foot}}\ and\ \bibinfo {author} {\bibfnamefont {S.}~\bibnamefont
  {Vagnozzi}},\ }\href {\doibase 10.1103/PhysRevD.91.023512} {\bibfield
  {journal} {\bibinfo  {journal} {Phys.Rev.}\ }\textbf {\bibinfo {volume}
  {D91}},\ \bibinfo {pages} {023512} (\bibinfo {year} {2015})},\ \Eprint
  {http://arxiv.org/abs/1409.7174} {arXiv:1409.7174 [hep-ph]} \BibitemShut
  {NoStop}%
\bibitem [{\citenamefont {Foot}(2014)}]{Foot:2014mia}%
  \BibitemOpen
  \bibfield  {author} {\bibinfo {author} {\bibfnamefont {R.}~\bibnamefont
  {Foot}},\ }\href {\doibase 10.1142/S0217751X14300130} {\bibfield  {journal}
  {\bibinfo  {journal} {Int.J.Mod.Phys.}\ }\textbf {\bibinfo {volume} {A29}},\
  \bibinfo {pages} {1430013} (\bibinfo {year} {2014})},\ \Eprint
  {http://arxiv.org/abs/1401.3965} {arXiv:1401.3965 [astro-ph.CO]} \BibitemShut
  {NoStop}%
\bibitem [{\citenamefont {Tulin}\ \emph {et~al.}(2013)\citenamefont {Tulin},
  \citenamefont {Yu},\ and\ \citenamefont {Zurek}}]{Tulin:2013teo}%
  \BibitemOpen
  \bibfield  {author} {\bibinfo {author} {\bibfnamefont {S.}~\bibnamefont
  {Tulin}}, \bibinfo {author} {\bibfnamefont {H.-B.}\ \bibnamefont {Yu}}, \
  and\ \bibinfo {author} {\bibfnamefont {K.~M.}\ \bibnamefont {Zurek}},\ }\href
  {\doibase 10.1103/PhysRevD.87.115007} {\bibfield  {journal} {\bibinfo
  {journal} {Phys.Rev.}\ }\textbf {\bibinfo {volume} {D87}},\ \bibinfo {pages}
  {115007} (\bibinfo {year} {2013})},\ \Eprint {http://arxiv.org/abs/1302.3898}
  {arXiv:1302.3898 [hep-ph]} \BibitemShut {NoStop}%
\bibitem [{\citenamefont {Cline}\ \emph
  {et~al.}(2014{\natexlab{b}})\citenamefont {Cline}, \citenamefont {Liu},
  \citenamefont {Moore},\ and\ \citenamefont {Xue}}]{Cline:2013zca}%
  \BibitemOpen
  \bibfield  {author} {\bibinfo {author} {\bibfnamefont {J.~M.}\ \bibnamefont
  {Cline}}, \bibinfo {author} {\bibfnamefont {Z.}~\bibnamefont {Liu}}, \bibinfo
  {author} {\bibfnamefont {G.}~\bibnamefont {Moore}}, \ and\ \bibinfo {author}
  {\bibfnamefont {W.}~\bibnamefont {Xue}},\ }\href {\doibase
  10.1103/PhysRevD.90.015023} {\bibfield  {journal} {\bibinfo  {journal}
  {Phys.Rev.}\ }\textbf {\bibinfo {volume} {D90}},\ \bibinfo {pages} {015023}
  (\bibinfo {year} {2014}{\natexlab{b}})},\ \Eprint
  {http://arxiv.org/abs/1312.3325} {arXiv:1312.3325 [hep-ph]} \BibitemShut
  {NoStop}%
\bibitem [{\citenamefont {Kouvaris}(2013)}]{Kouvaris:2013gya}%
  \BibitemOpen
  \bibfield  {author} {\bibinfo {author} {\bibfnamefont {C.}~\bibnamefont
  {Kouvaris}},\ }\href@noop {} {\bibfield  {journal} {\bibinfo  {journal}
  {Phys.Rev.}\ }\textbf {\bibinfo {volume} {D88}},\ \bibinfo {pages} {015001}
  (\bibinfo {year} {2013})},\ \Eprint {http://arxiv.org/abs/1304.7476}
  {arXiv:1304.7476 [hep-ph]} \BibitemShut {NoStop}%
\bibitem [{\citenamefont {Boddy}\ \emph
  {et~al.}(2014{\natexlab{a}})\citenamefont {Boddy}, \citenamefont {Feng},
  \citenamefont {Kaplinghat},\ and\ \citenamefont {Tait}}]{Boddy:2014yra}%
  \BibitemOpen
  \bibfield  {author} {\bibinfo {author} {\bibfnamefont {K.~K.}\ \bibnamefont
  {Boddy}}, \bibinfo {author} {\bibfnamefont {J.~L.}\ \bibnamefont {Feng}},
  \bibinfo {author} {\bibfnamefont {M.}~\bibnamefont {Kaplinghat}}, \ and\
  \bibinfo {author} {\bibfnamefont {T.~M.~P.}\ \bibnamefont {Tait}},\ }\href
  {\doibase 10.1103/PhysRevD.89.115017} {\bibfield  {journal} {\bibinfo
  {journal} {Phys.Rev.}\ }\textbf {\bibinfo {volume} {D89}},\ \bibinfo {pages}
  {115017} (\bibinfo {year} {2014}{\natexlab{a}})},\ \Eprint
  {http://arxiv.org/abs/1402.3629} {arXiv:1402.3629 [hep-ph]} \BibitemShut
  {NoStop}%
\bibitem [{\citenamefont {Hochberg}\ \emph {et~al.}(2014)\citenamefont
  {Hochberg}, \citenamefont {Kuflik}, \citenamefont {Volansky},\ and\
  \citenamefont {Wacker}}]{Hochberg:2014dra}%
  \BibitemOpen
  \bibfield  {author} {\bibinfo {author} {\bibfnamefont {Y.}~\bibnamefont
  {Hochberg}}, \bibinfo {author} {\bibfnamefont {E.}~\bibnamefont {Kuflik}},
  \bibinfo {author} {\bibfnamefont {T.}~\bibnamefont {Volansky}}, \ and\
  \bibinfo {author} {\bibfnamefont {J.~G.}\ \bibnamefont {Wacker}},\ }\href
  {\doibase 10.1103/PhysRevLett.113.171301} {\bibfield  {journal} {\bibinfo
  {journal} {Phys.Rev.Lett.}\ }\textbf {\bibinfo {volume} {113}},\ \bibinfo
  {pages} {171301} (\bibinfo {year} {2014})},\ \Eprint
  {http://arxiv.org/abs/1402.5143} {arXiv:1402.5143 [hep-ph]} \BibitemShut
  {NoStop}%
\bibitem [{\citenamefont {Kouvaris}\ \emph {et~al.}(2014)\citenamefont
  {Kouvaris}, \citenamefont {Shoemaker},\ and\ \citenamefont
  {Tuominen}}]{Kouvaris:2014uoa}%
  \BibitemOpen
  \bibfield  {author} {\bibinfo {author} {\bibfnamefont {C.}~\bibnamefont
  {Kouvaris}}, \bibinfo {author} {\bibfnamefont {I.~M.}\ \bibnamefont
  {Shoemaker}}, \ and\ \bibinfo {author} {\bibfnamefont {K.}~\bibnamefont
  {Tuominen}},\ }\href@noop {} {\  (\bibinfo {year} {2014})},\ \Eprint
  {http://arxiv.org/abs/1411.3730} {arXiv:1411.3730 [hep-ph]} \BibitemShut
  {NoStop}%
\bibitem [{\citenamefont {Weinberg}\ \emph {et~al.}(2013)\citenamefont
  {Weinberg}, \citenamefont {Bullock}, \citenamefont {Governato}, \citenamefont
  {de~Naray},\ and\ \citenamefont {Peter}}]{Weinberg:2013aya}%
  \BibitemOpen
  \bibfield  {author} {\bibinfo {author} {\bibfnamefont {D.~H.}\ \bibnamefont
  {Weinberg}}, \bibinfo {author} {\bibfnamefont {J.~S.}\ \bibnamefont
  {Bullock}}, \bibinfo {author} {\bibfnamefont {F.}~\bibnamefont {Governato}},
  \bibinfo {author} {\bibfnamefont {R.~K.}\ \bibnamefont {de~Naray}}, \ and\
  \bibinfo {author} {\bibfnamefont {A.~H.~G.}\ \bibnamefont {Peter}},\
  }\href@noop {} {\  (\bibinfo {year} {2013})},\ \Eprint
  {http://arxiv.org/abs/1306.0913} {arXiv:1306.0913 [astro-ph.CO]} \BibitemShut
  {NoStop}%
\bibitem [{\citenamefont {Pearce}\ and\ \citenamefont
  {Kusenko}(2013)}]{Pearce:2013ola}%
  \BibitemOpen
  \bibfield  {author} {\bibinfo {author} {\bibfnamefont {L.}~\bibnamefont
  {Pearce}}\ and\ \bibinfo {author} {\bibfnamefont {A.}~\bibnamefont
  {Kusenko}},\ }\href {\doibase 10.1103/PhysRevD.87.123531} {\bibfield
  {journal} {\bibinfo  {journal} {Phys.Rev.}\ }\textbf {\bibinfo {volume}
  {D87}},\ \bibinfo {pages} {123531} (\bibinfo {year} {2013})},\ \Eprint
  {http://arxiv.org/abs/1303.7294} {arXiv:1303.7294 [hep-ph]} \BibitemShut
  {NoStop}%
\bibitem [{\citenamefont {Frandsen}\ \emph {et~al.}(2014)\citenamefont
  {Frandsen}, \citenamefont {Sannino}, \citenamefont {Shoemaker},\ and\
  \citenamefont {Svendsen}}]{Frandsen:2014lfa}%
  \BibitemOpen
  \bibfield  {author} {\bibinfo {author} {\bibfnamefont {M.~T.}\ \bibnamefont
  {Frandsen}}, \bibinfo {author} {\bibfnamefont {F.}~\bibnamefont {Sannino}},
  \bibinfo {author} {\bibfnamefont {I.~M.}\ \bibnamefont {Shoemaker}}, \ and\
  \bibinfo {author} {\bibfnamefont {O.}~\bibnamefont {Svendsen}},\ }\href
  {\doibase 10.1088/1475-7516/2014/05/033} {\bibfield  {journal} {\bibinfo
  {journal} {JCAP}\ }\textbf {\bibinfo {volume} {1405}},\ \bibinfo {pages}
  {033} (\bibinfo {year} {2014})},\ \Eprint {http://arxiv.org/abs/1403.1570}
  {arXiv:1403.1570 [hep-ph]} \BibitemShut {NoStop}%
\bibitem [{\citenamefont {Cline}\ \emph
  {et~al.}(2014{\natexlab{c}})\citenamefont {Cline}, \citenamefont {Farzan},
  \citenamefont {Liu}, \citenamefont {Moore},\ and\ \citenamefont
  {Xue}}]{Cline:2014eaa}%
  \BibitemOpen
  \bibfield  {author} {\bibinfo {author} {\bibfnamefont {J.~M.}\ \bibnamefont
  {Cline}}, \bibinfo {author} {\bibfnamefont {Y.}~\bibnamefont {Farzan}},
  \bibinfo {author} {\bibfnamefont {Z.}~\bibnamefont {Liu}}, \bibinfo {author}
  {\bibfnamefont {G.~D.}\ \bibnamefont {Moore}}, \ and\ \bibinfo {author}
  {\bibfnamefont {W.}~\bibnamefont {Xue}},\ }\href {\doibase
  10.1103/PhysRevD.89.121302} {\bibfield  {journal} {\bibinfo  {journal}
  {Phys.Rev.}\ }\textbf {\bibinfo {volume} {D89}},\ \bibinfo {pages} {121302}
  (\bibinfo {year} {2014}{\natexlab{c}})},\ \Eprint
  {http://arxiv.org/abs/1404.3729} {arXiv:1404.3729 [hep-ph]} \BibitemShut
  {NoStop}%
\bibitem [{\citenamefont {Boddy}\ \emph
  {et~al.}(2014{\natexlab{b}})\citenamefont {Boddy}, \citenamefont {Feng},
  \citenamefont {Kaplinghat}, \citenamefont {Shadmi},\ and\ \citenamefont
  {Tait}}]{Boddy:2014qxa}%
  \BibitemOpen
  \bibfield  {author} {\bibinfo {author} {\bibfnamefont {K.~K.}\ \bibnamefont
  {Boddy}}, \bibinfo {author} {\bibfnamefont {J.~L.}\ \bibnamefont {Feng}},
  \bibinfo {author} {\bibfnamefont {M.}~\bibnamefont {Kaplinghat}}, \bibinfo
  {author} {\bibfnamefont {Y.}~\bibnamefont {Shadmi}}, \ and\ \bibinfo {author}
  {\bibfnamefont {T.~M.~P.}\ \bibnamefont {Tait}},\ }\href {\doibase
  10.1103/PhysRevD.90.095016} {\bibfield  {journal} {\bibinfo  {journal}
  {Phys.Rev.}\ }\textbf {\bibinfo {volume} {D90}},\ \bibinfo {pages} {095016}
  (\bibinfo {year} {2014}{\natexlab{b}})},\ \Eprint
  {http://arxiv.org/abs/1408.6532} {arXiv:1408.6532 [hep-ph]} \BibitemShut
  {NoStop}%
\bibitem [{\citenamefont {Detmold}\ \emph {et~al.}(2014)\citenamefont
  {Detmold}, \citenamefont {McCullough},\ and\ \citenamefont
  {Pochinsky}}]{Detmold:2014qqa}%
  \BibitemOpen
  \bibfield  {author} {\bibinfo {author} {\bibfnamefont {W.}~\bibnamefont
  {Detmold}}, \bibinfo {author} {\bibfnamefont {M.}~\bibnamefont {McCullough}},
  \ and\ \bibinfo {author} {\bibfnamefont {A.}~\bibnamefont {Pochinsky}},\
  }\href {\doibase 10.1103/PhysRevD.90.115013} {\bibfield  {journal} {\bibinfo
  {journal} {Phys.Rev.}\ }\textbf {\bibinfo {volume} {D90}},\ \bibinfo {pages}
  {115013} (\bibinfo {year} {2014})},\ \Eprint {http://arxiv.org/abs/1406.2276}
  {arXiv:1406.2276 [hep-ph]} \BibitemShut {NoStop}%
\bibitem [{\citenamefont {Pospelov}\ and\ \citenamefont
  {Ritz}(2009)}]{Pospelov:2008jd}%
  \BibitemOpen
  \bibfield  {author} {\bibinfo {author} {\bibfnamefont {M.}~\bibnamefont
  {Pospelov}}\ and\ \bibinfo {author} {\bibfnamefont {A.}~\bibnamefont
  {Ritz}},\ }\href {\doibase 10.1016/j.physletb.2008.12.012} {\bibfield
  {journal} {\bibinfo  {journal} {Phys.Lett.}\ }\textbf {\bibinfo {volume}
  {B671}},\ \bibinfo {pages} {391} (\bibinfo {year} {2009})},\ \Eprint
  {http://arxiv.org/abs/0810.1502} {arXiv:0810.1502 [hep-ph]} \BibitemShut
  {NoStop}%
\bibitem [{\citenamefont {March-Russell}\ and\ \citenamefont
  {West}(2009)}]{MarchRussell:2008tu}%
  \BibitemOpen
  \bibfield  {author} {\bibinfo {author} {\bibfnamefont {J.~D.}\ \bibnamefont
  {March-Russell}}\ and\ \bibinfo {author} {\bibfnamefont {S.~M.}\ \bibnamefont
  {West}},\ }\href {\doibase 10.1016/j.physletb.2009.04.010} {\bibfield
  {journal} {\bibinfo  {journal} {Phys.Lett.}\ }\textbf {\bibinfo {volume}
  {B676}},\ \bibinfo {pages} {133} (\bibinfo {year} {2009})},\ \Eprint
  {http://arxiv.org/abs/0812.0559} {arXiv:0812.0559 [astro-ph]} \BibitemShut
  {NoStop}%
\bibitem [{\citenamefont {Shepherd}\ \emph {et~al.}(2009)\citenamefont
  {Shepherd}, \citenamefont {Tait},\ and\ \citenamefont
  {Zaharijasmarch}}]{Shepherd:2009sa}%
  \BibitemOpen
  \bibfield  {author} {\bibinfo {author} {\bibfnamefont {W.}~\bibnamefont
  {Shepherd}}, \bibinfo {author} {\bibfnamefont {T.~M.}\ \bibnamefont {Tait}},
  \ and\ \bibinfo {author} {\bibfnamefont {G.}~\bibnamefont {Zaharijasmarch}},\
  }\href {\doibase 10.1103/PhysRevD.79.055022} {\bibfield  {journal} {\bibinfo
  {journal} {Phys.Rev.}\ }\textbf {\bibinfo {volume} {D79}},\ \bibinfo {pages}
  {055022} (\bibinfo {year} {2009})},\ \Eprint {http://arxiv.org/abs/0901.2125}
  {arXiv:0901.2125 [hep-ph]} \BibitemShut {NoStop}%
\bibitem [{\citenamefont {Graesser}\ \emph {et~al.}(2011)\citenamefont
  {Graesser}, \citenamefont {Shoemaker},\ and\ \citenamefont
  {Vecchi}}]{Graesser:2011wi}%
  \BibitemOpen
  \bibfield  {author} {\bibinfo {author} {\bibfnamefont {M.~L.}\ \bibnamefont
  {Graesser}}, \bibinfo {author} {\bibfnamefont {I.~M.}\ \bibnamefont
  {Shoemaker}}, \ and\ \bibinfo {author} {\bibfnamefont {L.}~\bibnamefont
  {Vecchi}},\ }\href {\doibase 10.1007/JHEP10(2011)110} {\bibfield  {journal}
  {\bibinfo  {journal} {JHEP}\ }\textbf {\bibinfo {volume} {1110}},\ \bibinfo
  {pages} {110} (\bibinfo {year} {2011})},\ \Eprint
  {http://arxiv.org/abs/1103.2771} {arXiv:1103.2771 [hep-ph]} \BibitemShut
  {NoStop}%
\bibitem [{\citenamefont {von Harling}\ and\ \citenamefont
  {Petraki}(2014)}]{vonHarling:2014kha}%
  \BibitemOpen
  \bibfield  {author} {\bibinfo {author} {\bibfnamefont {B.}~\bibnamefont {von
  Harling}}\ and\ \bibinfo {author} {\bibfnamefont {K.}~\bibnamefont
  {Petraki}},\ }\href {\doibase 10.1088/1475-7516/2014/12/033} {\bibfield
  {journal} {\bibinfo  {journal} {JCAP}\ }\textbf {\bibinfo {volume} {12}},\
  \bibinfo {pages} {033} (\bibinfo {year} {2014})},\ \Eprint
  {http://arxiv.org/abs/1407.7874} {arXiv:1407.7874 [hep-ph]} \BibitemShut
  {NoStop}%
\bibitem [{\citenamefont {{Akhiezer}}\ and\ \citenamefont
  {{Merenkov}}(1996)}]{AkhiezerMerenkov_sigmaHydrogen}%
  \BibitemOpen
  \bibfield  {author} {\bibinfo {author} {\bibfnamefont {A.~I.}\ \bibnamefont
  {{Akhiezer}}}\ and\ \bibinfo {author} {\bibfnamefont {N.~P.}\ \bibnamefont
  {{Merenkov}}},\ }\href {\doibase 10.1088/0953-4075} {\bibfield  {journal}
  {\bibinfo  {journal} {Journal of Physics B Atomic Molecular Physics}\
  }\textbf {\bibinfo {volume} {29}},\ \bibinfo {pages} {2135} (\bibinfo {year}
  {1996})}\BibitemShut {NoStop}%
\bibitem [{\citenamefont {Griest}\ and\ \citenamefont
  {Seckel}(1987)}]{Griest:1986yu}%
  \BibitemOpen
  \bibfield  {author} {\bibinfo {author} {\bibfnamefont {K.}~\bibnamefont
  {Griest}}\ and\ \bibinfo {author} {\bibfnamefont {D.}~\bibnamefont
  {Seckel}},\ }\href {\doibase 10.1016/0550-3213(87)90293-8} {\bibfield
  {journal} {\bibinfo  {journal} {Nucl.Phys.}\ }\textbf {\bibinfo {volume}
  {B283}},\ \bibinfo {pages} {681} (\bibinfo {year} {1987})}\BibitemShut
  {NoStop}%
\bibitem [{\citenamefont {Holdom}(1986)}]{Holdom:1985ag}%
  \BibitemOpen
  \bibfield  {author} {\bibinfo {author} {\bibfnamefont {B.}~\bibnamefont
  {Holdom}},\ }\href {\doibase 10.1016/0370-2693(86)91377-8} {\bibfield
  {journal} {\bibinfo  {journal} {Phys.Lett.}\ }\textbf {\bibinfo {volume}
  {B166}},\ \bibinfo {pages} {196} (\bibinfo {year} {1986})}\BibitemShut
  {NoStop}%
\bibitem [{\citenamefont {Foot}\ and\ \citenamefont {He}(1991)}]{Foot:1991kb}%
  \BibitemOpen
  \bibfield  {author} {\bibinfo {author} {\bibfnamefont {R.}~\bibnamefont
  {Foot}}\ and\ \bibinfo {author} {\bibfnamefont {X.-G.}\ \bibnamefont {He}},\
  }\href {\doibase 10.1016/0370-2693(91)90901-2} {\bibfield  {journal}
  {\bibinfo  {journal} {Phys.Lett.}\ }\textbf {\bibinfo {volume} {B267}},\
  \bibinfo {pages} {509} (\bibinfo {year} {1991})}\BibitemShut {NoStop}%
\bibitem [{\citenamefont {Batell}\ \emph {et~al.}(2009)\citenamefont {Batell},
  \citenamefont {Pospelov},\ and\ \citenamefont {Ritz}}]{Batell:2009yf}%
  \BibitemOpen
  \bibfield  {author} {\bibinfo {author} {\bibfnamefont {B.}~\bibnamefont
  {Batell}}, \bibinfo {author} {\bibfnamefont {M.}~\bibnamefont {Pospelov}}, \
  and\ \bibinfo {author} {\bibfnamefont {A.}~\bibnamefont {Ritz}},\ }\href
  {\doibase 10.1103/PhysRevD.79.115008} {\bibfield  {journal} {\bibinfo
  {journal} {Phys.Rev.}\ }\textbf {\bibinfo {volume} {D79}},\ \bibinfo {pages}
  {115008} (\bibinfo {year} {2009})},\ \Eprint {http://arxiv.org/abs/0903.0363}
  {arXiv:0903.0363 [hep-ph]} \BibitemShut {NoStop}%
\bibitem [{\citenamefont {Pospelov}\ \emph
  {et~al.}(2008{\natexlab{a}})\citenamefont {Pospelov}, \citenamefont {Ritz},\
  and\ \citenamefont {Voloshin}}]{Pospelov:2008jk}%
  \BibitemOpen
  \bibfield  {author} {\bibinfo {author} {\bibfnamefont {M.}~\bibnamefont
  {Pospelov}}, \bibinfo {author} {\bibfnamefont {A.}~\bibnamefont {Ritz}}, \
  and\ \bibinfo {author} {\bibfnamefont {M.~B.}\ \bibnamefont {Voloshin}},\
  }\href {\doibase 10.1103/PhysRevD.78.115012} {\bibfield  {journal} {\bibinfo
  {journal} {Phys.Rev.}\ }\textbf {\bibinfo {volume} {D78}},\ \bibinfo {pages}
  {115012} (\bibinfo {year} {2008}{\natexlab{a}})},\ \Eprint
  {http://arxiv.org/abs/0807.3279} {arXiv:0807.3279 [hep-ph]} \BibitemShut
  {NoStop}%
\bibitem [{\citenamefont {Jaeckel}\ \emph {et~al.}(2013)\citenamefont
  {Jaeckel}, \citenamefont {Jankowiak},\ and\ \citenamefont
  {Spannowsky}}]{Jaeckel:2012yz}%
  \BibitemOpen
  \bibfield  {author} {\bibinfo {author} {\bibfnamefont {J.}~\bibnamefont
  {Jaeckel}}, \bibinfo {author} {\bibfnamefont {M.}~\bibnamefont {Jankowiak}},
  \ and\ \bibinfo {author} {\bibfnamefont {M.}~\bibnamefont {Spannowsky}},\
  }\href {\doibase 10.1016/j.dark.2013.06.001} {\bibfield  {journal} {\bibinfo
  {journal} {Phys.Dark Univ.}\ }\textbf {\bibinfo {volume} {2}},\ \bibinfo
  {pages} {111} (\bibinfo {year} {2013})},\ \Eprint
  {http://arxiv.org/abs/1212.3620} {arXiv:1212.3620 [hep-ph]} \BibitemShut
  {NoStop}%
\bibitem [{\citenamefont {Lees}\ \emph {et~al.}(2014)\citenamefont {Lees} \emph
  {et~al.}}]{Lees:2014xha}%
  \BibitemOpen
  \bibfield  {author} {\bibinfo {author} {\bibfnamefont {J.}~\bibnamefont
  {Lees}} \emph {et~al.} (\bibinfo {collaboration} {BaBar Collaboration}),\
  }\href {\doibase 10.1103/PhysRevLett.113.201801} {\bibfield  {journal}
  {\bibinfo  {journal} {Phys.Rev.Lett.}\ }\textbf {\bibinfo {volume} {113}},\
  \bibinfo {pages} {201801} (\bibinfo {year} {2014})},\ \Eprint
  {http://arxiv.org/abs/1406.2980} {arXiv:1406.2980 [hep-ex]} \BibitemShut
  {NoStop}%
\bibitem [{\citenamefont {Hardy}\ \emph {et~al.}(2014)\citenamefont {Hardy},
  \citenamefont {Lasenby},\ and\ \citenamefont {Unwin}}]{Hardy:2014dea}%
  \BibitemOpen
  \bibfield  {author} {\bibinfo {author} {\bibfnamefont {E.}~\bibnamefont
  {Hardy}}, \bibinfo {author} {\bibfnamefont {R.}~\bibnamefont {Lasenby}}, \
  and\ \bibinfo {author} {\bibfnamefont {J.}~\bibnamefont {Unwin}},\ }\href
  {\doibase 10.1007/JHEP07(2014)049} {\bibfield  {journal} {\bibinfo  {journal}
  {JHEP}\ }\textbf {\bibinfo {volume} {1407}},\ \bibinfo {pages} {049}
  (\bibinfo {year} {2014})},\ \Eprint {http://arxiv.org/abs/1402.4500}
  {arXiv:1402.4500 [hep-ph]} \BibitemShut {NoStop}%
\bibitem [{\citenamefont {Bell}\ \emph {et~al.}(2015)\citenamefont {Bell},
  \citenamefont {Horiuchi},\ and\ \citenamefont {Shoemaker}}]{Bell:2014xta}%
  \BibitemOpen
  \bibfield  {author} {\bibinfo {author} {\bibfnamefont {N.~F.}\ \bibnamefont
  {Bell}}, \bibinfo {author} {\bibfnamefont {S.}~\bibnamefont {Horiuchi}}, \
  and\ \bibinfo {author} {\bibfnamefont {I.~M.}\ \bibnamefont {Shoemaker}},\
  }\href {\doibase 10.1103/PhysRevD.91.023505} {\bibfield  {journal} {\bibinfo
  {journal} {Phys.Rev.}\ }\textbf {\bibinfo {volume} {D91}},\ \bibinfo {pages}
  {023505} (\bibinfo {year} {2015})},\ \Eprint {http://arxiv.org/abs/1408.5142}
  {arXiv:1408.5142 [hep-ph]} \BibitemShut {NoStop}%
\bibitem [{\citenamefont {Cirelli}\ \emph {et~al.}(2012)\citenamefont
  {Cirelli}, \citenamefont {Panci}, \citenamefont {Servant},\ and\
  \citenamefont {Zaharijas}}]{Cirelli:2011ac}%
  \BibitemOpen
  \bibfield  {author} {\bibinfo {author} {\bibfnamefont {M.}~\bibnamefont
  {Cirelli}}, \bibinfo {author} {\bibfnamefont {P.}~\bibnamefont {Panci}},
  \bibinfo {author} {\bibfnamefont {G.}~\bibnamefont {Servant}}, \ and\
  \bibinfo {author} {\bibfnamefont {G.}~\bibnamefont {Zaharijas}},\ }\href
  {\doibase 10.1088/1475-7516} {\bibfield  {journal} {\bibinfo  {journal}
  {JCAP}\ }\textbf {\bibinfo {volume} {1203}},\ \bibinfo {pages} {015}
  (\bibinfo {year} {2012})},\ \Eprint {http://arxiv.org/abs/1110.3809}
  {arXiv:1110.3809 [hep-ph]} \BibitemShut {NoStop}%
\bibitem [{\citenamefont {Pospelov}\ \emph
  {et~al.}(2008{\natexlab{b}})\citenamefont {Pospelov}, \citenamefont {Ritz},\
  and\ \citenamefont {Voloshin}}]{Pospelov:2007mp}%
  \BibitemOpen
  \bibfield  {author} {\bibinfo {author} {\bibfnamefont {M.}~\bibnamefont
  {Pospelov}}, \bibinfo {author} {\bibfnamefont {A.}~\bibnamefont {Ritz}}, \
  and\ \bibinfo {author} {\bibfnamefont {M.~B.}\ \bibnamefont {Voloshin}},\
  }\href {\doibase 10.1016/j.physletb.2008.02.052} {\bibfield  {journal}
  {\bibinfo  {journal} {Phys.Lett.}\ }\textbf {\bibinfo {volume} {B662}},\
  \bibinfo {pages} {53} (\bibinfo {year} {2008}{\natexlab{b}})},\ \Eprint
  {http://arxiv.org/abs/0711.4866} {arXiv:0711.4866 [hep-ph]} \BibitemShut
  {NoStop}%
\bibitem [{\citenamefont {Steigman}\ \emph {et~al.}(2012)\citenamefont
  {Steigman}, \citenamefont {Dasgupta},\ and\ \citenamefont
  {Beacom}}]{Steigman:2012nb}%
  \BibitemOpen
  \bibfield  {author} {\bibinfo {author} {\bibfnamefont {G.}~\bibnamefont
  {Steigman}}, \bibinfo {author} {\bibfnamefont {B.}~\bibnamefont {Dasgupta}},
  \ and\ \bibinfo {author} {\bibfnamefont {J.~F.}\ \bibnamefont {Beacom}},\
  }\href {\doibase 10.1103/PhysRevD.86.023506} {\bibfield  {journal} {\bibinfo
  {journal} {Phys.Rev.}\ }\textbf {\bibinfo {volume} {D86}},\ \bibinfo {pages}
  {023506} (\bibinfo {year} {2012})},\ \Eprint {http://arxiv.org/abs/1204.3622}
  {arXiv:1204.3622 [hep-ph]} \BibitemShut {NoStop}%
\bibitem [{\citenamefont {Arkani-Hamed}\ \emph {et~al.}(2009)\citenamefont
  {Arkani-Hamed}, \citenamefont {Finkbeiner}, \citenamefont {Slatyer},\ and\
  \citenamefont {Weiner}}]{ArkaniHamed:2008qn}%
  \BibitemOpen
  \bibfield  {author} {\bibinfo {author} {\bibfnamefont {N.}~\bibnamefont
  {Arkani-Hamed}}, \bibinfo {author} {\bibfnamefont {D.~P.}\ \bibnamefont
  {Finkbeiner}}, \bibinfo {author} {\bibfnamefont {T.~R.}\ \bibnamefont
  {Slatyer}}, \ and\ \bibinfo {author} {\bibfnamefont {N.}~\bibnamefont
  {Weiner}},\ }\href {\doibase 10.1103/PhysRevD.79.015014} {\bibfield
  {journal} {\bibinfo  {journal} {Phys.Rev.}\ }\textbf {\bibinfo {volume}
  {D79}},\ \bibinfo {pages} {015014} (\bibinfo {year} {2009})},\ \Eprint
  {http://arxiv.org/abs/0810.0713} {arXiv:0810.0713 [hep-ph]} \BibitemShut
  {NoStop}%
\bibitem [{\citenamefont {Knodlseder}\ \emph {et~al.}(2005)\citenamefont
  {Knodlseder} \emph {et~al.}}]{Knodlseder:2005yq}%
  \BibitemOpen
  \bibfield  {author} {\bibinfo {author} {\bibfnamefont {J.}~\bibnamefont
  {Knodlseder}} \emph {et~al.},\ }\href {\doibase 10.1051/0004-6361:20042063}
  {\bibfield  {journal} {\bibinfo  {journal} {Astron. Astrophys.}\ }\textbf
  {\bibinfo {volume} {441}},\ \bibinfo {pages} {513} (\bibinfo {year}
  {2005})},\ \Eprint {http://arxiv.org/abs/astro-ph/0506026}
  {arXiv:astro-ph/0506026} \BibitemShut {NoStop}%
\bibitem [{\citenamefont {Beacom}\ and\ \citenamefont
  {Yuksel}(2006)}]{Beacom:2005qv}%
  \BibitemOpen
  \bibfield  {author} {\bibinfo {author} {\bibfnamefont {J.~F.}\ \bibnamefont
  {Beacom}}\ and\ \bibinfo {author} {\bibfnamefont {H.}~\bibnamefont
  {Yuksel}},\ }\href {\doibase 10.1103/PhysRevLett.97.071102} {\bibfield
  {journal} {\bibinfo  {journal} {Phys.Rev.Lett.}\ }\textbf {\bibinfo {volume}
  {97}},\ \bibinfo {pages} {071102} (\bibinfo {year} {2006})},\ \Eprint
  {http://arxiv.org/abs/astro-ph/0512411} {arXiv:astro-ph/0512411} \BibitemShut
  {NoStop}%
\bibitem [{\citenamefont {Bell}\ \emph {et~al.}(2010)\citenamefont {Bell},
  \citenamefont {Galea},\ and\ \citenamefont {Petraki}}]{Bell:2010fk}%
  \BibitemOpen
  \bibfield  {author} {\bibinfo {author} {\bibfnamefont {N.~F.}\ \bibnamefont
  {Bell}}, \bibinfo {author} {\bibfnamefont {A.~J.}\ \bibnamefont {Galea}}, \
  and\ \bibinfo {author} {\bibfnamefont {K.}~\bibnamefont {Petraki}},\ }\href
  {\doibase 10.1103/PhysRevD.82.023514} {\bibfield  {journal} {\bibinfo
  {journal} {Phys.Rev.}\ }\textbf {\bibinfo {volume} {D82}},\ \bibinfo {pages}
  {023514} (\bibinfo {year} {2010})},\ \Eprint {http://arxiv.org/abs/1004.1008}
  {arXiv:1004.1008 [astro-ph.HE]} \BibitemShut {NoStop}%
\bibitem [{\citenamefont {Beacom}\ \emph {et~al.}(2005)\citenamefont {Beacom},
  \citenamefont {Bell},\ and\ \citenamefont {Bertone}}]{Beacom:2004pe}%
  \BibitemOpen
  \bibfield  {author} {\bibinfo {author} {\bibfnamefont {J.~F.}\ \bibnamefont
  {Beacom}}, \bibinfo {author} {\bibfnamefont {N.~F.}\ \bibnamefont {Bell}}, \
  and\ \bibinfo {author} {\bibfnamefont {G.}~\bibnamefont {Bertone}},\ }\href
  {\doibase 10.1103/PhysRevLett.94.171301} {\bibfield  {journal} {\bibinfo
  {journal} {Phys.Rev.Lett.}\ }\textbf {\bibinfo {volume} {94}},\ \bibinfo
  {pages} {171301} (\bibinfo {year} {2005})},\ \Eprint
  {http://arxiv.org/abs/astro-ph/0409403} {arXiv:astro-ph/0409403} \BibitemShut
  {NoStop}%
\bibitem [{\citenamefont {Casse}\ \emph {et~al.}(2004)\citenamefont {Casse},
  \citenamefont {Cordier}, \citenamefont {Paul},\ and\ \citenamefont
  {Schanne}}]{Casse:2003fh}%
  \BibitemOpen
  \bibfield  {author} {\bibinfo {author} {\bibfnamefont {M.}~\bibnamefont
  {Casse}}, \bibinfo {author} {\bibfnamefont {B.}~\bibnamefont {Cordier}},
  \bibinfo {author} {\bibfnamefont {J.}~\bibnamefont {Paul}}, \ and\ \bibinfo
  {author} {\bibfnamefont {S.}~\bibnamefont {Schanne}},\ }\href {\doibase
  10.1086/381884} {\bibfield  {journal} {\bibinfo  {journal} {Astrophys.J.}\
  }\textbf {\bibinfo {volume} {602}},\ \bibinfo {pages} {L17} (\bibinfo {year}
  {2004})},\ \Eprint {http://arxiv.org/abs/astro-ph/0309824}
  {arXiv:astro-ph/0309824 [astro-ph]} \BibitemShut {NoStop}%
\bibitem [{\citenamefont {Bertone}\ \emph {et~al.}(2006)\citenamefont
  {Bertone}, \citenamefont {Kusenko}, \citenamefont {Palomares-Ruiz},
  \citenamefont {Pascoli},\ and\ \citenamefont {Semikoz}}]{Bertone:2004ek}%
  \BibitemOpen
  \bibfield  {author} {\bibinfo {author} {\bibfnamefont {G.}~\bibnamefont
  {Bertone}}, \bibinfo {author} {\bibfnamefont {A.}~\bibnamefont {Kusenko}},
  \bibinfo {author} {\bibfnamefont {S.}~\bibnamefont {Palomares-Ruiz}},
  \bibinfo {author} {\bibfnamefont {S.}~\bibnamefont {Pascoli}}, \ and\
  \bibinfo {author} {\bibfnamefont {D.}~\bibnamefont {Semikoz}},\ }\href
  {\doibase 10.1016/j.physletb.2006.03.022} {\bibfield  {journal} {\bibinfo
  {journal} {Phys.Lett.}\ }\textbf {\bibinfo {volume} {B636}},\ \bibinfo
  {pages} {20} (\bibinfo {year} {2006})},\ \Eprint
  {http://arxiv.org/abs/astro-ph/0405005} {arXiv:astro-ph/0405005 [astro-ph]}
  \BibitemShut {NoStop}%
\bibitem [{\citenamefont {Parizot}\ \emph {et~al.}(2005)\citenamefont
  {Parizot}, \citenamefont {Casse}, \citenamefont {Lehoucq},\ and\
  \citenamefont {Paul}}]{Parizot:2004ph}%
  \BibitemOpen
  \bibfield  {author} {\bibinfo {author} {\bibfnamefont {E.}~\bibnamefont
  {Parizot}}, \bibinfo {author} {\bibfnamefont {M.}~\bibnamefont {Casse}},
  \bibinfo {author} {\bibfnamefont {R.}~\bibnamefont {Lehoucq}}, \ and\
  \bibinfo {author} {\bibfnamefont {J.}~\bibnamefont {Paul}},\ }\href {\doibase
  10.1051/0004-6361:20042215} {\bibfield  {journal} {\bibinfo  {journal}
  {Astron.Astrophys.}\ }\textbf {\bibinfo {volume} {432}},\ \bibinfo {pages}
  {889} (\bibinfo {year} {2005})},\ \Eprint
  {http://arxiv.org/abs/astro-ph/0411656} {arXiv:astro-ph/0411656 [astro-ph]}
  \BibitemShut {NoStop}%
\bibitem [{\citenamefont {Prantzos}\ \emph {et~al.}(2011)\citenamefont
  {Prantzos}, \citenamefont {Boehm}, \citenamefont {Bykov}, \citenamefont
  {Diehl}, \citenamefont {Ferriere} \emph {et~al.}}]{Prantzos:2010wi}%
  \BibitemOpen
  \bibfield  {author} {\bibinfo {author} {\bibfnamefont {N.}~\bibnamefont
  {Prantzos}}, \bibinfo {author} {\bibfnamefont {C.}~\bibnamefont {Boehm}},
  \bibinfo {author} {\bibfnamefont {A.}~\bibnamefont {Bykov}}, \bibinfo
  {author} {\bibfnamefont {R.}~\bibnamefont {Diehl}}, \bibinfo {author}
  {\bibfnamefont {K.}~\bibnamefont {Ferriere}},  \emph {et~al.},\ }\href
  {\doibase 10.1103/RevModPhys.83.1001} {\bibfield  {journal} {\bibinfo
  {journal} {Rev.Mod.Phys.}\ }\textbf {\bibinfo {volume} {83}},\ \bibinfo
  {pages} {1001–1056} (\bibinfo {year} {2011})},\ \Eprint
  {http://arxiv.org/abs/1009.4620} {arXiv:1009.4620 [astro-ph.HE]} \BibitemShut
  {NoStop}%
\bibitem [{\citenamefont {Finkbeiner}\ and\ \citenamefont
  {Weiner}(2007)}]{Finkbeiner:2007kk}%
  \BibitemOpen
  \bibfield  {author} {\bibinfo {author} {\bibfnamefont {D.~P.}\ \bibnamefont
  {Finkbeiner}}\ and\ \bibinfo {author} {\bibfnamefont {N.}~\bibnamefont
  {Weiner}},\ }\href {\doibase 10.1103/PhysRevD.76.083519} {\bibfield
  {journal} {\bibinfo  {journal} {Phys.Rev.}\ }\textbf {\bibinfo {volume}
  {D76}},\ \bibinfo {pages} {083519} (\bibinfo {year} {2007})},\ \Eprint
  {http://arxiv.org/abs/astro-ph/0702587} {arXiv:astro-ph/0702587} \BibitemShut
  {NoStop}%
\bibitem [{\citenamefont {Gondolo}\ and\ \citenamefont
  {Silk}(1999)}]{Gondolo:1999ef}%
  \BibitemOpen
  \bibfield  {author} {\bibinfo {author} {\bibfnamefont {P.}~\bibnamefont
  {Gondolo}}\ and\ \bibinfo {author} {\bibfnamefont {J.}~\bibnamefont {Silk}},\
  }\href {\doibase 10.1103/PhysRevLett.83.1719} {\bibfield  {journal} {\bibinfo
   {journal} {Phys.Rev.Lett.}\ }\textbf {\bibinfo {volume} {83}},\ \bibinfo
  {pages} {1719} (\bibinfo {year} {1999})},\ \Eprint
  {http://arxiv.org/abs/astro-ph/9906391} {arXiv:astro-ph/9906391 [astro-ph]}
  \BibitemShut {NoStop}%
\bibitem [{\citenamefont {Blumenthal}\ \emph {et~al.}(1986)\citenamefont
  {Blumenthal}, \citenamefont {Faber}, \citenamefont {Flores},\ and\
  \citenamefont {Primack}}]{Blumenthal:1985qy}%
  \BibitemOpen
  \bibfield  {author} {\bibinfo {author} {\bibfnamefont {G.~R.}\ \bibnamefont
  {Blumenthal}}, \bibinfo {author} {\bibfnamefont {S.}~\bibnamefont {Faber}},
  \bibinfo {author} {\bibfnamefont {R.}~\bibnamefont {Flores}}, \ and\ \bibinfo
  {author} {\bibfnamefont {J.~R.}\ \bibnamefont {Primack}},\ }\href {\doibase
  10.1086/163867} {\bibfield  {journal} {\bibinfo  {journal} {Astrophys.J.}\
  }\textbf {\bibinfo {volume} {301}},\ \bibinfo {pages} {27} (\bibinfo {year}
  {1986})}\BibitemShut {NoStop}%
\bibitem [{\citenamefont {Gnedin}\ and\ \citenamefont
  {Primack}(2004)}]{Gnedin:2003rj}%
  \BibitemOpen
  \bibfield  {author} {\bibinfo {author} {\bibfnamefont {O.~Y.}\ \bibnamefont
  {Gnedin}}\ and\ \bibinfo {author} {\bibfnamefont {J.~R.}\ \bibnamefont
  {Primack}},\ }\href {\doibase 10.1103/PhysRevLett.93.061302} {\bibfield
  {journal} {\bibinfo  {journal} {Phys.Rev.Lett.}\ }\textbf {\bibinfo {volume}
  {93}},\ \bibinfo {pages} {061302} (\bibinfo {year} {2004})},\ \Eprint
  {http://arxiv.org/abs/astro-ph/0308385} {arXiv:astro-ph/0308385 [astro-ph]}
  \BibitemShut {NoStop}%
\bibitem [{\citenamefont {Gustafsson}\ \emph {et~al.}(2006)\citenamefont
  {Gustafsson}, \citenamefont {Fairbairn},\ and\ \citenamefont
  {Sommer-Larsen}}]{Gustafsson:2006gr}%
  \BibitemOpen
  \bibfield  {author} {\bibinfo {author} {\bibfnamefont {M.}~\bibnamefont
  {Gustafsson}}, \bibinfo {author} {\bibfnamefont {M.}~\bibnamefont
  {Fairbairn}}, \ and\ \bibinfo {author} {\bibfnamefont {J.}~\bibnamefont
  {Sommer-Larsen}},\ }\href {\doibase 10.1103/PhysRevD.74.123522} {\bibfield
  {journal} {\bibinfo  {journal} {Phys.Rev.}\ }\textbf {\bibinfo {volume}
  {D74}},\ \bibinfo {pages} {123522} (\bibinfo {year} {2006})},\ \Eprint
  {http://arxiv.org/abs/astro-ph/0608634} {arXiv:astro-ph/0608634 [astro-ph]}
  \BibitemShut {NoStop}%
\bibitem [{\citenamefont {Gnedin}\ \emph {et~al.}(2011)\citenamefont {Gnedin},
  \citenamefont {Ceverino}, \citenamefont {Gnedin}, \citenamefont {Klypin},
  \citenamefont {Kravtsov} \emph {et~al.}}]{Gnedin:2011uj}%
  \BibitemOpen
  \bibfield  {author} {\bibinfo {author} {\bibfnamefont {O.~Y.}\ \bibnamefont
  {Gnedin}}, \bibinfo {author} {\bibfnamefont {D.}~\bibnamefont {Ceverino}},
  \bibinfo {author} {\bibfnamefont {N.~Y.}\ \bibnamefont {Gnedin}}, \bibinfo
  {author} {\bibfnamefont {A.~A.}\ \bibnamefont {Klypin}}, \bibinfo {author}
  {\bibfnamefont {A.~V.}\ \bibnamefont {Kravtsov}},  \emph {et~al.},\
  }\href@noop {} {\  (\bibinfo {year} {2011})},\ \Eprint
  {http://arxiv.org/abs/1108.5736} {arXiv:1108.5736 [astro-ph.CO]} \BibitemShut
  {NoStop}%
\bibitem [{\citenamefont {Madhavacheril}\ \emph {et~al.}(2014)\citenamefont
  {Madhavacheril}, \citenamefont {Sehgal},\ and\ \citenamefont
  {Slatyer}}]{Madhavacheril:2013cna}%
  \BibitemOpen
  \bibfield  {author} {\bibinfo {author} {\bibfnamefont {M.~S.}\ \bibnamefont
  {Madhavacheril}}, \bibinfo {author} {\bibfnamefont {N.}~\bibnamefont
  {Sehgal}}, \ and\ \bibinfo {author} {\bibfnamefont {T.~R.}\ \bibnamefont
  {Slatyer}},\ }\href {\doibase 10.1103/PhysRevD.89.103508} {\bibfield
  {journal} {\bibinfo  {journal} {Phys.Rev.}\ }\textbf {\bibinfo {volume}
  {D89}},\ \bibinfo {pages} {103508} (\bibinfo {year} {2014})},\ \Eprint
  {http://arxiv.org/abs/1310.3815} {arXiv:1310.3815 [astro-ph.CO]} \BibitemShut
  {NoStop}%
\bibitem [{\citenamefont {Fornengo}\ \emph {et~al.}(2011)\citenamefont
  {Fornengo}, \citenamefont {Panci},\ and\ \citenamefont
  {Regis}}]{Fornengo:2011sz}%
  \BibitemOpen
  \bibfield  {author} {\bibinfo {author} {\bibfnamefont {N.}~\bibnamefont
  {Fornengo}}, \bibinfo {author} {\bibfnamefont {P.}~\bibnamefont {Panci}}, \
  and\ \bibinfo {author} {\bibfnamefont {M.}~\bibnamefont {Regis}},\ }\href
  {\doibase 10.1103/PhysRevD.84.115002} {\bibfield  {journal} {\bibinfo
  {journal} {Phys.Rev.}\ }\textbf {\bibinfo {volume} {D84}},\ \bibinfo {pages}
  {115002} (\bibinfo {year} {2011})},\ \Eprint {http://arxiv.org/abs/1108.4661}
  {arXiv:1108.4661 [hep-ph]} \BibitemShut {NoStop}%
\bibitem [{\citenamefont {Kaplinghat}\ \emph {et~al.}(2014)\citenamefont
  {Kaplinghat}, \citenamefont {Tulin},\ and\ \citenamefont
  {Yu}}]{Kaplinghat:2013yxa}%
  \BibitemOpen
  \bibfield  {author} {\bibinfo {author} {\bibfnamefont {M.}~\bibnamefont
  {Kaplinghat}}, \bibinfo {author} {\bibfnamefont {S.}~\bibnamefont {Tulin}}, \
  and\ \bibinfo {author} {\bibfnamefont {H.-B.}\ \bibnamefont {Yu}},\ }\href
  {\doibase 10.1103/PhysRevD.89.035009} {\bibfield  {journal} {\bibinfo
  {journal} {Phys.Rev.}\ }\textbf {\bibinfo {volume} {D89}},\ \bibinfo {pages}
  {035009} (\bibinfo {year} {2014})},\ \Eprint {http://arxiv.org/abs/1310.7945}
  {arXiv:1310.7945 [hep-ph]} \BibitemShut {NoStop}%
\bibitem [{\citenamefont {Akerib}\ \emph {et~al.}(2014)\citenamefont {Akerib}
  \emph {et~al.}}]{Akerib:2013tjd}%
  \BibitemOpen
  \bibfield  {author} {\bibinfo {author} {\bibfnamefont {D.}~\bibnamefont
  {Akerib}} \emph {et~al.} (\bibinfo {collaboration} {LUX Collaboration}),\
  }\href {\doibase 10.1103/PhysRevLett.112.091303} {\bibfield  {journal}
  {\bibinfo  {journal} {Phys.Rev.Lett.}\ }\textbf {\bibinfo {volume} {112}},\
  \bibinfo {pages} {091303} (\bibinfo {year} {2014})},\ \Eprint
  {http://arxiv.org/abs/1310.8214} {arXiv:1310.8214 [astro-ph.CO]} \BibitemShut
  {NoStop}%
\end{thebibliography}%
\end{document}